\documentclass[12pt]{article}
\usepackage[english]{babel}
\usepackage{amsmath,amssymb,dsfont, float, bm}
\usepackage{amsmath,amssymb,amsfonts,amsthm,mathtools,mathrsfs,booktabs}
\usepackage{multicol}
\usepackage{multirow}
\usepackage{tikz,float}
\usepackage{pgfplots}
\pgfplotsset{compat=1.16}
\usepackage{authblk}
\usepackage{caption}
\usepackage{graphicx}
\usepackage{subcaption}
\usepackage{rotating}

\usepackage{natbib}

\usepackage[colorlinks=true, citecolor=blue, urlcolor=blue]{hyperref}
\usepackage{newtxtext,newtxmath} 
\usepackage{multirow}
\usepackage{bookmark}
\usepackage{booktabs}
\hypersetup{                 
colorlinks=true,
linkcolor=blue,
citecolor=blue,
anchorcolor=blue,
urlcolor=blue,
linktoc=page,
}
\usepackage{pdflscape}
\usepackage{adjustbox}
\usepackage{chngcntr}
\usepackage{threeparttable}
\usepackage{caption}

\usepackage[a4paper,margin=2.5cm]{geometry}

\begin{document}

\title{\bf A modified score function for monotone likelihood in promotion time cure rate models}

\author[1]{Stephany Lima de Oliveira}
\author[2]{\, Frederico Machado
Almeida \thanks{Corresponding author: frederico.almeida@unb.br}}
\affil[1]{Centro de Gestão e Estudos Estratégicos, Bras\'ilia, Brazil.}
\affil[2]{Department of Statistics, University of Bras\'ilia, 70910-900, Bras\'ilia, Brazil.}
\setcounter{Maxaffil}{0}

\maketitle
\begin{abstract}
Survival models that incorporate a cure fraction provide a flexible framework for jointly modeling the cure and the survival distributions. However, when the data comprise a high proportion of censored observations or highly unbalanced binary covariates, maximum likelihood estimation may become unstable, leading to parameter estimates that diverge to $\pm\infty$. This phenomenon, commonly referred to in the literature as monotone likelihood (ML), compromises statistical inference by precluding the existence of finite maximum likelihood estimates. Specifically, the likelihood function increases monotonically along certain directions in the parameter space, so no finite maximizer exists. To the best of our knowledge, the ML problem has received little or no attention in the context involving the promotion time model. This paper addresses this gap by proposing a modified score function based on Firth's bias-reduction method, which adjusts the estimation procedure to ensure finite and stable parameter estimates. The performance of the proposed approach is evaluated through extensive Monte Carlo simulation studies. An application to a real dataset further demonstrates its practical advantages, showing that the mitosis factor, an established prognostic marker, becomes statistically significant under the proposed methodology.
\end{abstract}

\vspace{0.2cm}

\section{Introduction}\label{sec1}

Classical survival analysis techniques assume that all experimental units are susceptible to the event of interest; in other words, every individual will eventually experience the event if observed for a sufficient period. Consequently, the survival function is expected to approach zero as time approaches infinity. However, this assumption does not always hold. In medical studies, for example, advances in diagnostic methods and therapeutic interventions have enabled some patients to achieve a sustained response to treatment, making them no longer susceptible to the event of interest. Such individuals are commonly referred to as immune, cured, or long-term survivors \citep{berkson1952, yakovlev1996}.

Similar situations are observed in various other fields. In engineering, particularly within reliability studies, populations exhibiting limited failures may include a subset of components that remain operational throughout the entire follow-up period \citep{meeker1987}. In demography, some married couples may never experience divorce \citep{kulu2014marriage}. Similarly, in studies of criminal recidivism, a proportion of individuals released from prison may never re-offend \citep{schmidt1989predicting}. Collectively, these examples illustrate contexts in which a fraction of the population remains permanently unaffected by the event of interest.

The presence of long-term survivors challenges the validity of conventional survival models, as these models are not designed to account for immune individuals and may, consequently, yield biased estimates. To address this issue, two principal classes of cure rate models have been developed: the standard mixture cure model (SMCM) and the promotion time cure model (PTCM). The SMCM, initially introduced by \citet{boag1949} and subsequently extended by \citet{berkson1952}, posits that the study population comprises a mixture of two subgroups: cured and uncured individuals. In contrast, the PTCM, proposed by \citet{yakovlev1996} and serving as the primary focus of this work, was developed as an alternative modeling framework.

The PTCM is formulated by introducing a latent random variable that quantifies the number of latent competing causes associated with the occurrence of the event of interest. The literature further describes two unified frameworks that encompass both the SMCM and the PTCM. The first, proposed by \citet{ibrahim2005unified}, is based on the Box-Cox transformation. Subsequently, \citet{rodrigues2009unification} introduced a second unified approach, emphasizing that the latent variable typically follows a Bernoulli distribution under the standard mixture cure model, whereas a Poisson distribution is commonly assumed in the promotion time cure model.

As in conventional settings, model parameters are typically estimated using maximum likelihood. However, in many practical situations, particularly those involving rare events or a high proportion of censored observations, the likelihood function may fail to attain a finite maximum with respect to one or more model parameters. Consequently, point estimates, confidence intervals, and other likelihood-based inferential quantities may diverge toward $\pm\infty$. This phenomenon, commonly referred to as ML in the literature, arises directly from the configuration of the observed data.

In the context of survival analysis, the occurrence of ML may be exacerbated by a high proportion of censored observations, resulting in a pronounced imbalance between censored and observed failure times. Nevertheless, studies by \citet{heinze2001} and \citet{almeida2018} have demonstrated that ML is not solely attributable to excessive censoring; it may also arise from severely unbalanced binary covariates, particularly when all observed failures are confined to a single category of a binary predictor. These findings suggest that the risk of encountering ML increases with the number of binary covariates included in the regression model. Moreover, \citet{Klein2003} observed that ML can also occur in models containing only continuous covariates, provided these covariates display a strong monotonic association with the ordering of failure times.

The melanoma dataset originally analyzed by \citet{cherobin2018} served as the primary impetus for this study. In short, melanoma is among the most aggressive forms of skin cancer and is associated with considerable mortality, particularly when diagnosed at advanced stages. Efforts to reduce melanoma-related mortality center on three main strategies: $(1)$ early detection of primary melanoma; $(2)$ identification of patients at elevated risk for metastasis; and $(3)$ determination of prognostic factors linked to disease progression. A detailed description of the dataset is provided in Section~\ref{sec4}.

\begin{figure}[H]
\centering
\includegraphics[width=0.6\linewidth]{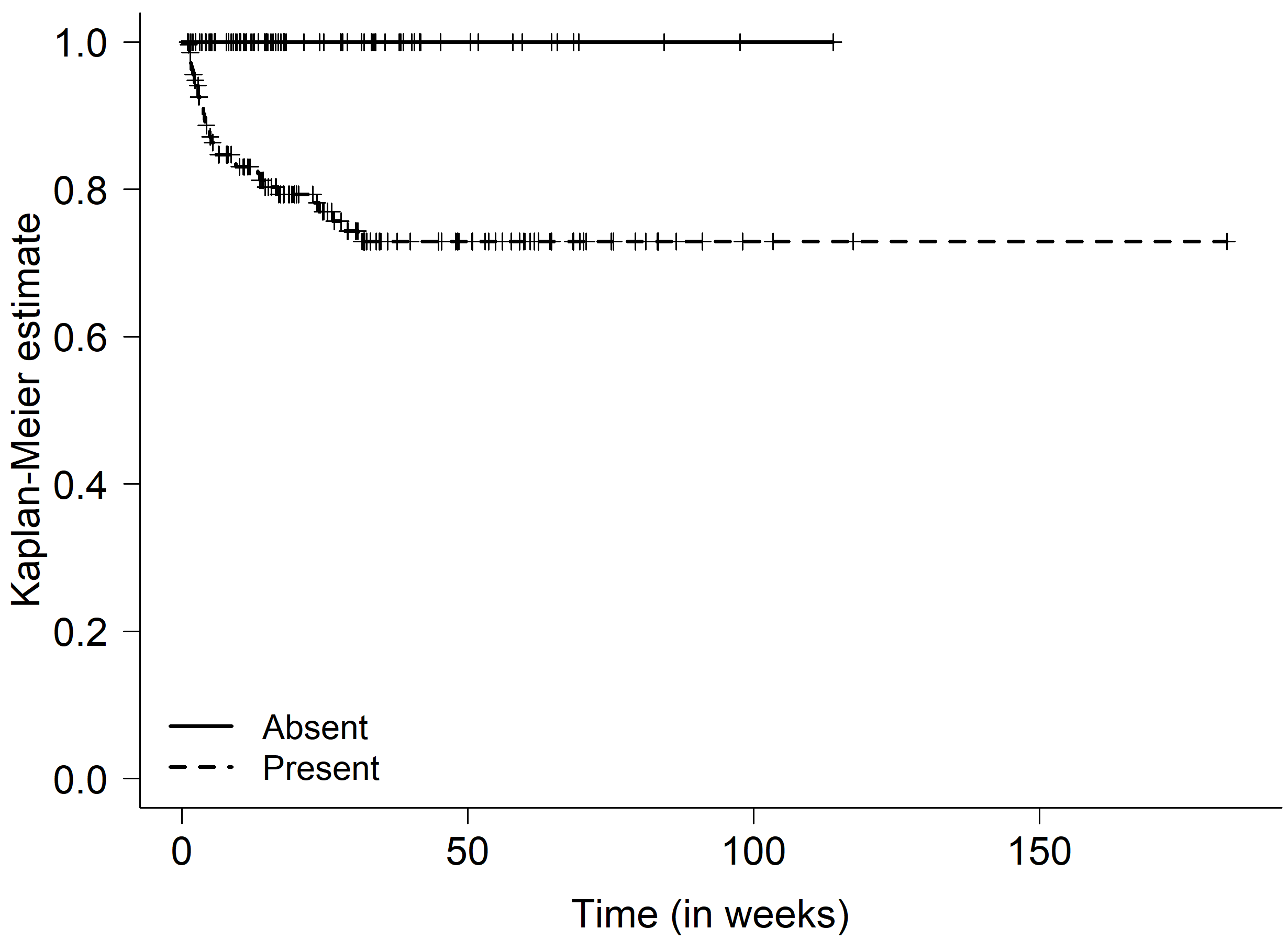}\\
\caption{Kaplan-Meier curve for the mitosis risk factor.}
\label{fig:EKMmitosis}
\end{figure}

The Kaplan-Meier (KM) estimator, introduced by \citet{kaplan1958}, serves as a valuable graphical tool for detecting ML and identifying evidence of a cured fraction in survival data. Since the KM estimator only changes at observed failure times, the absence of failures in one or more categories of a categorical covariate can clearly indicate ML. In these cases, the associated KM curve remains at a survival probability of one throughout the entire follow-up period. This pattern is illustrated in Figure~\ref{fig:EKMmitosis}, based on data from the melanoma dataset. For instance, no patients without mitosis experienced metastasis, the event of interest, indicating that all observed failures were restricted to patients with mitosis. This complete separation of failures across levels of the mitosis covariate exemplifies an ML issue in the melanoma dataset. Additionally, among patients with mitosis, the KM curve plateaus at a survival probability above 60\%, providing strong evidence for a substantial cured fraction in this subgroup.

Several strategies for addressing the ML problem have been proposed in the literature. These can be broadly categorized as remedial or non-remedial approaches. Among the remedial strategies, notable examples include: $(i)$ excluding the covariate responsible for ML and refitting the model without adjusting for its effect \citep{heinze2001}; $(ii)$ proceeding with inference despite the presence of ML and interpreting the resulting estimates, even when they are unstable or imprecise due to the underlying issue; $(iii)$ stratifying the variable associated with ML and fitting separate models within each stratum \citep{bryson1981incidence}; and $(iv)$ re-parameterizing the model or adopting alternative model specifications \citep{andersson1997, heinze2002separation}.

Additionally, several statistical methods have been developed to address the problem of ML. Unlike ad hoc or merely remedial strategies, these approaches directly address the underlying causes of ML, yielding finite, interpretable, and statistically efficient parameter estimates. The most prominent methods include: $(i)$ Firth’s bias-reduction technique \citep{firth1993, heinze2001}, one of the most widely applied frequentist solutions for ML; $(ii)$ Bayesian approaches, which incorporate prior distributions to stabilize and regularize parameter estimates \citep{greenland2007prior, lin2013shrinkage, almeida2023}; $(iii)$ penalized estimation techniques, such as Ridge and Lasso regression; and, where relevant, $(iv)$ data imputation methods to address missing covariate values \citep{morais2025}, among others.

The ML issue has been extensively examined across a broad array of statistical models, including generalized linear models for binary and categorical responses \citep{heinze2003fixing, Kholassa2016}, longitudinal data models \citep{Kenne2024bias}, and survival models \citep{heinze2001, almeida2018, almeida2022}, among others. Nevertheless, despite the well-documented prevalence of ML, particularly in medical research involving rare events or substantial censoring, there remains a notable scarcity of statistically principled and robust solutions specifically designed for cure rate models. Under the SMCM, only three studies have explicitly addressed the ML issue, namely, \citet{almeida2021}, \citet{almeida2022}, and \citet{almeida2023}. In contrast, methodological advancements about the PTCM remain virtually nonexistent, i.e., to the best of our knowledge, a comprehensive review of the scientific literature, using the keywords ``monotone likelihood'' and ``promotion time cure model'', did not identify any published studies specifically examining the ML issue in the context of PTCMs. 

Motivated by the apparent gap in the literature and recognizing the importance of ML in statistical modeling, this paper proposes: $(1)$ the development of a modified score function, based on Firth’s bias-reduction method, to ensure finite and reliable parameter estimates, even in the presence of ML; $(2)$ a comprehensive evaluation of the finite-sample performance of the proposed methodology through extensive Monte Carlo (MC) simulation studies; and $(3)$ an empirical demonstration of the practical importance of the proposed approach, using a real dataset characterized by both a cure fraction and the occurrence of ML issues.

The remainder of this paper is organized as follows. Section~\ref{sec2} presents the formulation of the PTM, describes the conventional estimation procedure, and develops the modified score function based on Firth’s bias-reduction method. Section~\ref{sec3} investigates the finite-sample properties of Firth's modified estimators through a comprehensive MC simulation study. Section~\ref{sec4} demonstrates the practical applicability of the proposed methodology using a real melanoma dataset. Finally, Section~\ref{sec5} summarizes the main findings and offers directions for future research.

\section{Model formulation}\label{sec2}

The PTM considered in this paper provides advantages for the biological interpretation of cure by modeling the event as arising from a latent number of independently acting competing risk factors. To illustrate the model construction, let $M$ denote the number of competing causes associated with the event of interest, where $M$ is a discrete random variable that is typically assumed to follow a Poisson distribution with mean $\theta>0$, i.e., $\mathbb{P}\left(M=m|\theta\right)=\theta^m\exp(-\theta)/m!$, with $m \in\mathbb{N}$. For each $j = 1, \ldots, M$, let $Z_j$ denote the latent promotion time until the $j$-th metastatic cell gives rise to detectable metastatic disease. Conditional on $M = m$, the random variables $Z_j$ are assumed to be independent and identically distributed with a common promotion cumulative distribution function $F(z) = 1 - S(z)$, which does not depend on $M$. The observed time to cancer relapse is defined as $T = \min\{Z_j: 1 \leq j \leq M\}$. Following \cite{ibrahim2005unified} and \cite{rodrigues2009unification}, the survival function for the overall population is defined as

\vspace{-0.5cm}
\begin{eqnarray}
S_{p}(t|\bm{\psi})&=& \mathbb{P}\left(M=0\right) + \mathbb{P}\left(Z_1>t,\cdots, Z_M>t, M \geq 1\right)\nonumber\\
   &=& \exp\left(-\theta\right) + \sum\limits_{m=1}^{\infty} S(t|\bm{\vartheta})^{m}\frac{\theta^{m}}{m!}\exp\left(-\theta\right)\nonumber\\
    &=& \exp\left(-\theta F(t|\bm{\vartheta})\right),
   \label{survPOP}
\end{eqnarray}

\noindent where $\bm{\psi} = (\theta, \bm{\vartheta}^{\top})^{\top}$ is the $d$-dimensional model parameter vector, and $\bm{\vartheta} = (\vartheta_1, \vartheta_2, \ldots, \vartheta_{q})^{\top}$ is the $q$-dimensional vector associated with the distribution for susceptible subjects.

It is important to note that when $M = 0$, the subject is considered non-susceptible to the event of interest, indicating the absence of competing causes. Consequently, $\mathbb{P}\left(Z_0 = \infty\right) = 1$, which implies that the survival time for long-term survivors is degenerate at infinity; equivalently, the survival function of $Z_0$ is identically equal to one. Additionally, the population survival function defined in~\eqref{survPOP} satisfies the following properties: $(i)$ $\lim\limits_{t \to 0^+} S_p(t|\bm{\psi}) = 1$; $(ii)$ $S_p(t|\bm{\psi})$ is a non-increasing function in $t$; and $(iii)$ $\mathbb{P}\left(M=0\right)= \lim\limits_{t \to +\infty} S_p(t|\bm{\psi})=\exp(-\theta)$. The last property shows that $S_p(t|\bm{\psi})$ is an improper (or defective) survival function. Furthermore, the associated population density and hazard functions are given by $f_p(t|\bm{\psi})=\theta f(t|\bm{\vartheta}) \exp\left(-\theta F(t|\bm{\vartheta})\right)$ and $h_p(t|\bm{\psi})=\frac{\partial \Lambda_p(t|\bm{\psi})}{\partial t} = \theta f(t|\bm{\vartheta})$, where $F(t|\bm{\vartheta}) = \Lambda_p(t|\bm{\psi})/\theta$ denotes the normalized cumulative distribution function. 

Despite its origins in biological applications, the model in~\eqref{survPOP} applies to a wide range of fields where the primary interest is in modeling the time until a particular event occurs \citep{bremhorst2019,redivo2024}. 

Assume that $\mathbf{x} = (x_0, x_1, \ldots, x_{q_1})^{\top}$ is the covariate vector influencing the cumulative distribution function $F(t|\bm{\vartheta})$ for the non-cured subjects, where $x_0 = 1$ represents the intercept. Let $\bm{\beta} = (\beta_0, \beta_1, \ldots, \beta_{q_1})^{\top} \in \bm{\vartheta}$ denote the associated regression coefficients, where $\bm{\beta}$ is defined in $\mathbb{R}^{q_1+1}$. Under the accelerated failure time model, the logarithm of the survival time associated with the covariate vector $\mathbf{x}$ is specified as $\log T_{\mathbf{x}} = \mu(\mathbf{x}) + \sigma W$. Thus, $\log T_{\mathbf{x}}$ follows a location-scale model with location parameter $\mu(\mathbf{x})=\mathbf{x}^{\top} \bm{\beta}$ and scale parameter $\sigma>0$. The error term $W$ can be expressed as

\begin{equation}
W = \frac{\log T_{\mathbf{x}}-\mu(\mathbf{x})}{\sigma}.
\label{error.term}
\end{equation}

\noindent{Given their flexibility and successful application in previous studies, the random variable $W$ will be modeled using the standard logistic and standard extreme-value distributions, respectively.}

\subsection{Estimation procedure}\label{subsc1}

Assume that $D_{oi} = \left\{(t_i, \delta_i, \mathbf{x}_i): i = 1, \ldots, n\right\}$ denotes the observed data, where, for each $i$, the observed survival time is given by $t_i = \min\left(T_i, C_i\right)$. Here, $T_i$ and $C_i$ are independent random variables representing the failure and censoring times, respectively. The failure time $T_i$ is defined as $T_i = \min\left(Z_{i1}, Z_{i2}, \ldots, Z_{iM_i}\right)$, where $Z_{i1}, Z_{i2}, \ldots, Z_{iM_i}$ denote latent failure times corresponding to different competing risks, and $Z_{i0}$ is a constant variable degenerate $+\infty$. Additionally, the failure indicator is defined as $\delta_i = \mathds{1}_{\{T_i \leq C_i\}}$, where $\mathds{1}_{\mathcal{B}}(y_i)$ denotes the indicator function. The $q$-dimensional parameter set associated with the latency distribution is $\bm{\vartheta} = (\alpha,\bm{\beta}^{\top})^{\top}$, where $\alpha>0$ is the shape parameter of the distribution of $T_i$, and $\bm{\beta}$ is the $(q_1+1)$-dimensional regression coefficient vector. Let $\mathbf{x}_i=\left(x_{i0}, x_{i1}, x_{i2}, \ldots, x_{ip}\right)^{\top}$ the $i$-th row of the full rank design matrix $\mathbf{X} \in \mathbb{R}^{n \times (q_1+1)}$, where $x_{i0} = 1$ for all $i = 1, 2, \ldots, n$.

Under the non-informative right censoring mechanism, the observed-data likelihood function is given by:

\vspace{-0.3cm}
\begin{equation}
\label{ObsLikelihood}
L(\bm{\psi}|D_{\text{o}})= \prod_{i=1}^{n} f_{\text{p}}(t_i| {\bf x}_i,\bm{\psi})^{\delta_i} \, S_{\text{p}}(t_i|{\bf x}_i,\bm{\psi})^{1 - \delta_i}.
\end{equation}

\noindent From Equation~\eqref{ObsLikelihood}, the observed log-likelihood function is $\ell(\bm{\psi}) = \log L(\bm{\psi}|D_{\mathrm{o}})$. In this case, the maximum likelihood estimator (MLE) of the vector $\bm{\psi}$ can be obtained as $\hat{\bm{\psi}} = \underset{\bm{\psi} \in \mathbb{R}^d}{\sup}\,\ell(\bm{\psi})$. Equivalently, the MLE $\bm{\hat{\psi}}$ of $\bm{\psi}$ can be obtained by solving the likelihood equation:

\begin{equation}
\mathrm{U}\left(\boldsymbol{\psi}\right) = \mathbf{0}_d,
\label{usualScore}
\end{equation}

\noindent where $\mathrm{U}\left(\bm{\psi}\right) = \frac{\partial \ell(\bm{\psi})}{\partial \bm{\psi}}$ denotes the standard score function and $\mathbf{0}_d$ is a $d$-dimensional vector of zeros. Under the conditions described in Section~\ref{sec1}, maximizing $\ell\left(\bm{\psi}\right)$ can be particularly challenging when ML is present or when the data are highly unbalanced. In such cases, finite values for the MLEs may not be obtained; that is, one or more components $\psi_k \in \bm{\psi}$, $k = 1, 2, \ldots, d$, can diverge to infinity. Equivalently, the corresponding profile score equation may have no finite root, so that $\mathrm{U}(\psi_k) \neq 0$ for some $k$.

Among the various approaches proposed in the literature to address ML, as reviewed in Section~\ref{sec1}, we propose a strategy that modifies the score function with an adjustment term based on the mean bias-reduction technique, which has proven effective for estimation problems of this kind \citep{kosmid2020}.

\subsection{Modified score function}\label{subsc2}

Firth's bias-reduction method was originally proposed by \citet{firth1993} to reduce the first-order bias of MLEs in generalized linear models. Its widespread adoption is largely due to adaptations by \citet{heinze2001,heinze2002separation}, who demonstrated that this penalization can ensure finite parameter estimates even in the presence of ML, also known as separation. More recently, \citet{almeida2021} applied Firth's method to cure rate models, with particular emphasis on the SMCM. This implicit bias-reduction procedure incorporates an adjustment term, derived from the first-order bias approximation via a Taylor series expansion, into the score function. This modification yields finite parameter estimates even when the score equation in~\eqref{usualScore} has no finite solution due to an ML issue.

Without loss of generality, let $\bm{\psi} = (\psi_1,\ldots,\psi_d)^{\top}$ denote the vector of model parameters. For regular models, \citet{cox1974} showed that the asymptotic bias of the maximum likelihood estimator for the parameter $\psi_{\nu}$ can be expressed as,

\begin{equation}
b(\psi_u) = \mathbb{E}(\hat{\psi}_{u} - \psi_{u})
\approx \frac{b_1(\psi_u)}{n} + O(n^{-2}),
\label{biasExp}
\end{equation}

\noindent where $n$ denotes the sample size, $b_1(\psi_u)/n$ is the first-order asymptotic bias term, and $u \in \{1,2,\cdots,d\}$. The Firth bias-reduction method modifies the standard score function presented in Equation~\eqref{usualScore} by incorporating an adjustment term that eliminates the first-order bias. The resulting modified score function is given by

\begin{equation}
\mathbf{U}^*(\bm{\psi}) = \mathbf{U}(\bm{\psi}) + \mathbf{A}\left(\bm{\psi}\right),
\label{modScore}
\end{equation}

\noindent where $\mathbf{A}(\boldsymbol{\psi})$ serves as the adjustment term and can be expressed as $\mathbf{A}\left(\bm{\psi}\right)= -\mathbf{I}\left(\bm{\psi}\right)\mathbf{b}(\bm{\psi})$,
being $\mathbf{b}(\bm{\psi})$ the vector of first-order asymptotic bias terms obtained from Equation~\eqref{biasExp}, and $\mathbf{I}\left(\bm{\psi}\right)$ is the Fisher information matrix. Equivalently, each component $A\left(\psi_u\right) \in \mathbf{A}\left(\bm{\psi}\right)$, can be written as

\begin{equation}
    A(\psi_u) = \frac{1}{2}\sum\limits_{\nu=1}^{d}\sum\limits_{b=1}^{d}\mbox{I}^{\nu,b}\left(\bm{\psi}\right)\xi_{\nu,b,u}=
    \frac{1}{2}\mathrm{tr}\!\left[
        \mathbf{I}^{-1}(\bm{\psi}) \left(
        \frac{\partial \mathbf{I}(\bm{\psi})}{\partial \psi_u}
    \right)\right],
    \label{penalty}
\end{equation}

\noindent with $\xi_{\nu,b,u} = \frac{\partial \mathrm{I}^{\nu,b}\left(\bm{\psi}\right)}{\partial \psi_u}$. The quantities $\mathrm{I}^{\nu,b}(\bm{\psi})$ and $\mathrm{I}_{\nu,b}\left(\bm{\psi}\right)$ denote the $(\nu,b)$-th entries of $\mathrm{I}^{-1}(\bm{\psi})$ and $\mathrm{I}(\bm{\psi})$, respectively. Based on expression~\eqref{modScore}, the modified (or penalized) MLE, $\hat{\bm{\psi}}^*$, is obtained by solving the modified likelihood equation $\mathbf{U}^*(\bm{\psi}) = \mathbf{0}_d$, i.e., $\mathbf{U}\left(\bm{\psi}\right)=-\mathbf{A}\left(\bm{\psi}\right)$.

Therefore, after some algebraic manipulation, the penalized maximum likelihood estimator can equivalently be obtained by maximizing the following expression $\ell^*\left(\bm{\psi}\right) = \ell\left(\bm{\psi}\right) + \frac{1}{2}\log\left|\mathbf{I}(\boldsymbol{\psi})\right|$,
where $\left|\mathbf{I}(\boldsymbol{\psi})\right|$ denotes the determinant of the matrix $\mathbf{I}(\bm{\psi})$. Equivalently, the penalized likelihood function can be written as

\begin{equation}
L^*(\bm{\psi}) = L(\bm{\psi})\,|\mathbf{I}(\bm{\psi})|^{1/2}.
\label{penLik}
\end{equation}

\noindent For distributions in exponential families with canonical parametrization, Firth's method is equivalent to penalizing the likelihood with Jeffreys' invariant prior \citep{jeffreys1946}. When the expected information matrix is not available in closed form, it may be replaced by the observed information matrix, since
$n^{-1}\mathcal{H}\left(\hat{\bm{\psi}}^*\right) \overset{\mathrm{p}}{\longrightarrow} \mathbf{I}\left(\bm{\psi}\right)$. Thus, the resulting penalized maximum likelihood estimator can be interpreted as the posterior mode under Jeffreys' prior. Nevertheless, despite this formal connection between the penalized likelihood and the corresponding posterior distribution, the estimation procedure remains entirely frequentist.

Details of the derivation of the first-, second-, and third-order cumulants are provided in Appendix~\ref{apendiceA1}, where the expected information matrix ${\bf I}(\bm{\psi})$ is replaced by the observed information matrix $\mathcal{H}(\bm{\psi})$. Under the usual regularity conditions, \citet{firth1993} established the existence and uniqueness of the penalized maximum likelihood estimator, even in settings characterized by substantial data imbalance. Moreover, $\hat{\bm{\psi}}^{*}$ is consistent and asymptotically normal, that is, $n^{1/2}\left(\hat{\bm{\psi}}^{*}-\bm{\psi}\right)\overset{\mathcal{D}}{\longrightarrow}
\mathcal{N}_d\left(\mathbf{0},{\bf I}^{-1}\left(\bm{\psi}\right)\right)$.

\section{Simulation study}\label{sec3}

This section investigates the finite-sample performance of the proposed methodology. The quality of the modified MLEs is rigorously evaluated using MC simulation studies conducted under various scenarios, with particular emphasis on the two parametric models examined in this study.

For the MC simulations, two covariates were incorporated into the regression structure. The first is a binary variable configured as follows: the initial six observations are designated as successes (case group), $x_{1i}=1$, whereas the remaining $n$-6 observations are classified as failures (control group), $x_{1i}=0$, where $i\in \{1,2,\cdots, n\}$ and $n \in \{50, 300, 600, 1000\}$ denotes the sample sizes. This configuration is designed to control the proportion of samples exhibiting ML issues across the $G=1000$ MC replications. The second covariate is continuous and generated from a standard normal distribution. Accordingly, the covariate vector that influences the long-term survival distributions is denoted by $\mathbf{x}_i = \left(1, x_{1i}, x_{2i}\right)^\top$. 

The promotion survival times were generated as follows. First, a latent random variable $M_i\in\{0, 1, 2, \ldots \}$, representing the number of latent competing causes, was drawn from a Poisson distribution with mean $\theta$; that is, $M_i \sim \mathrm{Poisson}\left(\theta\right)$. If $M_i>0$, indicating the presence of at least one latent cause contributing to the event under study, $M_i$ latent survival times $Z_{im}$ (for $m=1, \ldots, M_i$) were generated using the inverse transform method, such that $Z_{im}=F^{-1}(u_{im})$, where $F(\cdot)$ denotes the baseline cumulative distribution function related to the survival times presented in~\eqref{error.term} and each $u_{im} \sim \mathcal{U}(0, 1)$ is an independent realization from the standard uniform distribution. The failure time for the $i$-th susceptible individual ($M_i>0$) was defined as $T_i = \min\{Z_{i1}, Z_{i2},\ldots, Z_{iM_i}\}$. If $M_i = 0$, indicating that the $i$-th individual is cured, the failure time was set to $T_i = Z_{i0}$, where $Z_{i0}$ is a constant degenerate e at infinity; that is, $Z_{i0} \equiv \infty$.

The censoring times $C_i$ were generated from an exponential distribution with mean $\mathbb{E}(C_i) = \tau$. To control the censoring and cure rates, as well as the proportion of MC samples exhibiting ML issue, $\tau$ was fixed at $1.50$. Accordingly, the observed survival time and failure indicator for the susceptible subpopulation are given by $t_i^* = \min(T_i, C_i)$ and $w_i = \mathds{1}_{\{T_i \leq C_i\}}$, respectively. To ensure model identifiability, the population-level survival time is defined as $t_i = t_i^*$ if $M_i > 0$ and $t_i = C_i$ if $M_i = 0$. The corresponding population-level censoring indicator is given by $\delta_i = w_i$ if $M_i > 0$, and $\delta_i = 0$ if $M_i = 0$.

Let $\psi_{b} \in \bm{\psi}$ denote the true value of the $b$-th model parameter, and $\hat{\psi}_{b}^{(j)}$ the modified MLE in the $j$-th MC sample, for $j = 1,2,\ldots,G$. The measures used to evaluate the estimation performance of the parameter are as follows: $(i)$ relative bias (Rbias), expressed as: $\mbox{Rbias}=(1/G)\sum\limits_{j=1}^G\left(\hat{\psi}_{b}^{*(j)}-\psi_{b}\right)/|\psi_{b}|$; $(ii)$ asymptotic standard error (ASE), defined as: $\mbox{ASE}=(1/G)\sum\limits_{j=1}^G\mbox{ASE}(\hat{\psi}_b^{*(j)})$; $(iii)$ empirical standard deviation (ESD), given by: $\mbox{ESD}=\left[\sum\limits_{j=1}^G\left(\hat{\psi}_{b}^{*(j)}-\bar{\hat{\psi}}_{b}^{*}\right)^2/(G-1)\right]^{1/2}$; $(iv)$ root mean square error (RMSE) computed in the following manner:  $\mbox{RMSE}=\left[(1/G)\sum\limits_{j=1}^G\left(\hat{\psi}_{b}^{*(j)}-\psi_{b}\right)^2\right]^{1/2}$ and $(v)$ coverage rate (CR) defined as: $\mbox{CR}=(1/G)\sum\limits_{j=1}^{G}\mathds{1}\left(\psi_b\,\in\, \mbox{CI}^{(j)}\left(\psi_b\right)\right)$. Here, $\mbox{CI}^{(j)}\left(\psi_b\right)=\hat{\psi}_{b}^{*(j)}\pm 1.96\,\mbox{SE}(\hat{\psi}_b^{*(j)})$ denotes the 95\% confidence interval for the parameter $\psi_b$ in the $j$-th MC sample, $\mbox{SE}(\hat{\psi}_b^{*(j)})=\sqrt{I^{b,b}\left(\hat{\bm{\psi}}^{*}\right)}$ is the corresponding asymptotics standard error and $\bar{\hat{\psi}}_{b}^{*}=(1/G)\sum\limits_{j=1}^G \hat{\psi}_{b}^{*(j)}$.

To define the scenarios analyzed in this study, a sensitivity analysis was conducted by considering different combinations of the parameters of interest. For simplicity, two scenarios were evaluated. In both cases, the censoring and cure rate proportions were fixed at 70\% and 40\%, respectively. Additionally, the approximate proportion of samples exhibiting ML issue was 40\% in Scenario 1 and 20\% in Scenario 2. The true parameter values for both models and scenarios are shown in Table~\ref{Truevalues}. All simulations were performed using R software \citep{RCoreTeam2026}.

\begin{table}[H]
\raggedleft
\begin{threeparttable}
\captionsetup{
 font={sf},
 justification=raggedright,
 singlelinecheck=false}
\caption{True parameter values employed in the simulation studies for the PTM-Weibull and PTM-Log-Logistic regression models.}
\vspace{-0.2cm}
\vspace{0.1cm}
\label{Truevalues}
\setlength{\tabcolsep}{0.2cm}
{\sf
\begin{tabular}{ccccccc}
\toprule
Scenario & $\theta$ & $\beta_0$ & $\beta_1$ & $\beta_2$ & $\alpha$ &  \% ML \\
\midrule
1 & 0.9 & $-$0.8 & 1.5 & 1.2 & 2.0 &  $\approx 40$ \\
2 & 0.9 & $-$0.8 & 0.8 & 1.2 & 1.5 & $\approx 20$ \\
\bottomrule
\end{tabular}
}
\end{threeparttable}
\end{table}

The Monte Carlo simulation results for the two regression models examined in this study, across both scenarios, are presented in Table~\ref{Simures}. The findings highlight the pronounced impact of the ML problem on the estimation of the coefficient $\beta_1$ associated with the binary and highly unbalanced covariate $x_{1i}$, particularly in the PTM-Weibull regression, where the relative bias remains substantial, approximately -22\% in Scenario 1 and -13\% in Scenario 2. Furthermore, the ESD consistently exceeds the ASE across all scenarios and in both models, indicating a substantial discrepancy between the empirical and asymptotic measures of estimation variability and, consequently, suggesting that the models yield unreliable estimates of the parameter of interest. Similarly, the coverage rate falls short of the nominal 95\% level in both scenarios and across all models examined.

\begin{table}[H]
	\centering
    \captionsetup{
    font={sf},
    justification=raggedright,
    singlelinecheck=false}
   \caption{Monte Carlo simulation study results for the Weibull and Log-Logistic PTM under Scenarios 1 and 2.}
    \vspace{-0.2cm}
	\label{Simures}
    \sf
    {\footnotesize
    \setlength{\tabcolsep}{4pt}
	  \begin{tabular}{crrrrrrrrrrr}
	\toprule
    \multicolumn{12}{c}{\normalsize \sf PTM-Weibull regression}\\
	\cline{1-12}
    \multirow{2}{*}{$n$} & \multirow{2}{*}{Measures}& \multicolumn{5}{c}{Scenario 1} & \multicolumn{5}{c}{Scenario 2} \\
	\cmidrule(lr){3-7}\cmidrule(lr){8-12}
	&  & $\theta$ & $\beta_0$ & $\beta_1$ & $\beta_2$ & $\alpha$ & $\theta$ & $\beta_0$ & $\beta_1$ & $\beta_2$ & $\alpha$ \\
   \midrule
	\multirow{6}{*}{50}
	& Rbias & -0.047 & -0.109 & -0.229 & -0.015 & 0.279 & 0.001 & -0.098 & -0.086 & -0.016 & 0.253 \\
	& ASE   & 0.251 & 0.163 & 0.464 & 0.164 & 0.484 & 0.264 & 0.226 & 0.549 & 0.219 & 0.345 \\
	& ESD  & 0.332 & 0.267 & 0.743 & 0.234 & 0.690 & 0.365 & 0.380 & 0.915 & 0.305 & 0.480 \\
	& RMSE  & 0.334 & 0.281 & 0.818 & 0.234 & 0.887 & 0.365 & 0.388 & 0.918 & 0.305 & 0.612 \\
	& CR    & 0.858 & 0.746 & 0.720 & 0.834 & 0.859 & 0.855 & 0.726 & 0.755 & 0.836 & 0.839 \\
	\midrule

	\multirow{6}{*}{300}
	& Rbias & -0.011 & -0.027 & -0.187 & -0.001 & 0.039 & -0.012 & -0.042 & -0.088 & -0.006 & 0.035 \\
	& ASE   & 0.101 & 0.078 & 0.499 & 0.072 & 0.158 & 0.103 & 0.108 & 0.602 & 0.094 & 0.118 \\
	& ESD  & 0.107 & 0.090 & 0.846 & 0.076 & 0.163 & 0.109 & 0.123 & 0.919 & 0.095 & 0.126 \\
	& RMSE  & 0.107 & 0.092 & 0.891 & 0.076 & 0.181 & 0.109 & 0.127 & 0.921 & 0.095 & 0.136 \\
	& CR  & 0.933 & 0.903 & 0.769 & 0.938 & 0.931 & 0.933 & 0.895 & 0.851 & 0.944 & 0.932 \\
	\midrule
   \multirow{6}{*}{600}
	& Rbias & -0.006 & -0.014 & -0.223 & -0.001 & 0.019 & -0.006 & -0.020 & -0.136 & -0.003 & 0.018 \\
	& ASE   & 0.072 & 0.057 & 0.508 & 0.051 & 0.109 & 0.073 & 0.078 & 0.609 & 0.067 & 0.083 \\
	& ESD  & 0.076 & 0.055 & 1.082 & 0.049 & 0.113 & 0.075 & 0.081 & 0.957 & 0.070 & 0.081 \\
	& RMSE  & 0.076 & 0.056 & 1.132 & 0.049 & 0.120 & 0.075 & 0.082 & 0.963 & 0.070 & 0.085 \\
	& CR    & 0.926 & 0.942 & 0.729 & 0.956 & 0.940 & 0.941 & 0.931 & 0.828 & 0.936 & 0.952 \\
	\midrule

	\multirow{6}{*}{1000}
	& Rbias & -0.003 & -0.008 & -0.204 & 0.001 & 0.013 & -0.003 & -0.014 & -0.195 & -0.001 & 0.011 \\
	& ASE   & 0.056 & 0.044 & 0.515 & 0.039 & 0.084 & 0.057 & 0.061 & 0.605 & 0.051 & 0.064 \\
	& ESD  & 0.057 & 0.047 & 0.867 & 0.041 & 0.088 & 0.058 & 0.065 & 0.899 & 0.052 & 0.062 \\
	& RMSE  & 0.057 & 0.047 & 0.919 & 0.041 & 0.092 & 0.058 & 0.066 & 0.912 & 0.052 & 0.064 \\
	& CR    & 0.940 & 0.924 & 0.785 & 0.936 & 0.930 & 0.943 & 0.935 & 0.842 & 0.950 & 0.950 \\
	\toprule
    \multicolumn{12}{c}{\normalsize \sf PTM-Log-Logistic regression}\\
	\cline{1-12}
    \multirow{2}{*}{$n$} & \multirow{2}{*}{Measures} & \multicolumn{5}{c}{Scenario 1} & \multicolumn{5}{c}{Scenario 2} \\
	\cmidrule(lr){3-7}\cmidrule(lr){8-12}
	&  & $\theta$ & $\beta_0$ & $\beta_1$ & $\beta_2$ & $\alpha$ & $\theta$ & $\beta_0$ & $\beta_1$ & $\beta_2$ & $\alpha$ \\
    \hline
    \multirow{5}{*}{50} 
	& Rbias&-0.027&-0.112&-0.116&-0.036& 0.322& 0.064&-0.115& 0.113&-0.053& 0.342\\
	& ASE  & 0.427& 0.414& 0.832& 0.300& 0.872& 0.345& 0.376& 0.702& 0.275& 0.465\\
	& ESD & 0.427& 0.414& 0.832& 0.300& 0.872& 0.571& 0.589& 1.067& 0.367& 0.672\\
	& RMSE & 0.428& 0.423& 0.850& 0.303& 1.083& 0.574& 0.596& 1.070& 0.373& 0.846\\
	& CR & 0.840& 0.767& 0.794& 0.846& 0.885& 0.807& 0.734& 0.793& 0.829& 0.860\\
	\midrule
	\multirow{5}{*}{300} 
	& Rbias&-0.013&-0.032&-0.090&-0.002& 0.045& 0.004&-0.043& 0.126&-0.014& 0.046\\
	& ASE  & 0.116& 0.130& 0.611& 0.097& 0.201& 0.124& 0.181& 0.752& 0.123& 0.156\\
	& ESD & 0.135& 0.152& 0.878& 0.104& 0.211& 0.197& 0.245& 1.025& 0.122& 0.170\\
	& RMSE & 0.136& 0.154& 0.888& 0.104& 0.230& 0.197& 0.247& 1.029& 0.123& 0.183\\
	& CR & 0.924& 0.891& 0.805& 0.939& 0.936& 0.870& 0.855& 0.886& 0.952& 0.927\\
	\midrule
	\multirow{5}{*}{600} 
	& Rbias&-0.014&-0.030&-0.158&-0.004& 0.028&-0.011&-0.030& 0.045&-0.005& 0.022\\
	& ASE  & 0.082& 0.094& 0.605& 0.069& 0.141& 0.089& 0.134& 0.760& 0.088& 0.109\\
	& ESD & 0.087& 0.097& 1.110& 0.068& 0.148& 0.091& 0.141& 1.175& 0.093& 0.108\\
	& RMSE & 0.088& 0.100& 1.134& 0.069& 0.158& 0.092& 0.143& 1.175& 0.094& 0.113\\
	& CR & 0.923& 0.920& 0.788& 0.945& 0.940& 0.937& 0.927& 0.886& 0.952& 0.927\\
	\midrule
	\multirow{5}{*}{1000} 
	& Rbias&-0.006&-0.012&-0.022& 0.000& 0.015&-0.009&-0.030& 0.134&-0.002& 0.017\\
	& ASE & 0.064& 0.074& 0.636& 0.054& 0.108& 0.069& 0.104& 0.760& 0.068& 0.084\\
	& ESD & 0.064& 0.076& 0.875& 0.056& 0.109& 0.071& 0.110& 1.021& 0.068& 0.085\\
	& RMSE & 0.064& 0.077& 0.875& 0.056& 0.114& 0.071& 0.113& 1.026& 0.068& 0.089\\
	& CR & 0.941& 0.936& 0.856& 0.945& 0.943& 0.935& 0.923& 0.834& 0.931& 0.953\\    
    \bottomrule
	\end{tabular}
     }
\end{table}

Although the relative bias decreases as the sample size increases, and the RMSE, ESD, and ASE display similar values, the estimates of the parameters $\theta$ and $\beta_0$, which are not associated with any covariate, remain negatively affected by the ML problem, particularly when the sample size is small. The coverage rates are slightly below the nominal level in both regression models and across all scenarios, with this shortfall being particularly evident for smaller sample sizes.

\noindent Similarly, the parameters $\beta_2$ and $\alpha$, which are not directly associated with the ML problem, exhibit noticeable estimation issues when $n=50$. However, the performance of the evaluated metrics improves rapidly as the sample size increases. For example, as the relative bias decreases toward zero, the precision measures (ESD, ASE, and RMSE) converge to similar values. Finally, the coverage probability stabilizes around its nominal level as the sample size grows, a pattern observed in both regression models and under both scenarios. Overall, these results suggest that the impact of the ML problem on parameters not directly associated with the highly imbalanced binary covariate is more evident in small samples. However, as the information in the data increases, the influence of the ML problem gradually diminishes. All of these findings are consistent with the histograms presented in Figure~\ref{Perform3} of Appendix~\ref{apendiceA2}.

It is worth emphasizing that the poorer performance of the measures associated with the parameter $\beta_1$ is likely related to the severe imbalance in the binary covariate. Notably, when $n=50$, the proportion of successes in $x_{1i}$, that is, $\%\{x_{1i}=1\}$, is approximately 12\%, corresponding to a ratio of 6 successes to 44 failures (6:44). However, given the configuration adopted in this study, this ratio shifts to 6:294 for $n=300$, 6:594 for $n=600$, and, under extreme imbalance, 6:994 for $n = 1000$. It is noteworthy that, in the latter case, the proportion of successes in the binary covariate is only $0.6\%$. Accordingly, as the sample size $n$ increases, the number of failures in $x_{1i}$ necessarily grows, while the number of successes remains constant. This increasingly pronounced imbalance may substantially reduce the precision of parameter estimates of the binary covariate, as reflected in the reported performance metrics.

To assess the validity of the preceding statement, a balancing experiment was conducted by considering different configurations of $x_{1i}$. The objective was to investigate the behavior of the performance measures for the parameters $\beta_0$, $\beta_1$, and $\theta$, with particular emphasis on $\beta_1$, especially in relation to the ML problem. The sample size was fixed at $n = 1000$, with the most severely imbalanced configuration of $x_{1i}$ represented by a success-to-failure ratio of 6:994. Thirteen distinct configurations of the binary covariate were examined. In the first configuration, the proportion of successes was set to $0.6\%$, corresponding to 6 successes and 994 failures. The number of successes in $x_{1i}$ was then progressively increased until a perfectly balanced configuration was reached, with a 50:50 success-to-failure ratio. The final configuration comprised $99.4\%$ successes, resulting in a ratio of 994:6.

The results of the balancing experiment are presented in Figure~\ref{fig:balanced} and Figures~\ref{Perform3} and~\ref{Perform4} of Appendix~\ref{apendiceA2}. These figures demonstrate that, for both models and under both scenarios, the performance of the evaluated measures improved when the proportion of $x_{1i}=1$ was between $7.5\%$ and $92.5\%$. Specifically, the relative bias stabilized around zero, the ASE and RMSE decreased rapidly, and the coverage probabilities converged to the nominal level of $95\%$. However, under highly imbalanced configurations, with $0.6\%$ and $99.4\%$ successes, these measures deteriorated, exhibiting larger values of relative bias, ASE, and RMSE. Moreover, the coverage probabilities in these extreme cases tended to fall below the nominal level.

\begin{figure}[H]
\centering
 \includegraphics[scale=0.4]{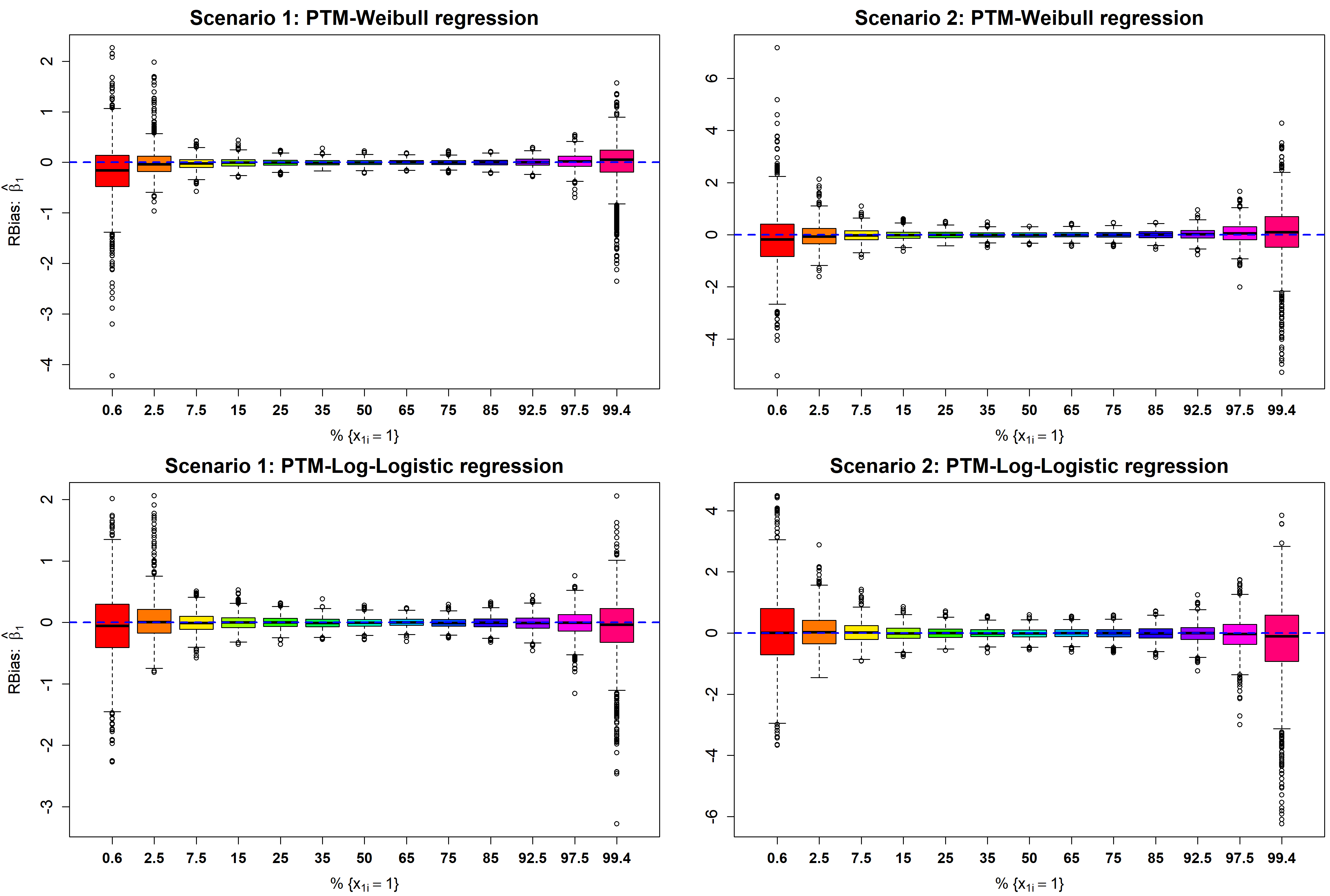}\\ 
 \includegraphics[scale=0.4]{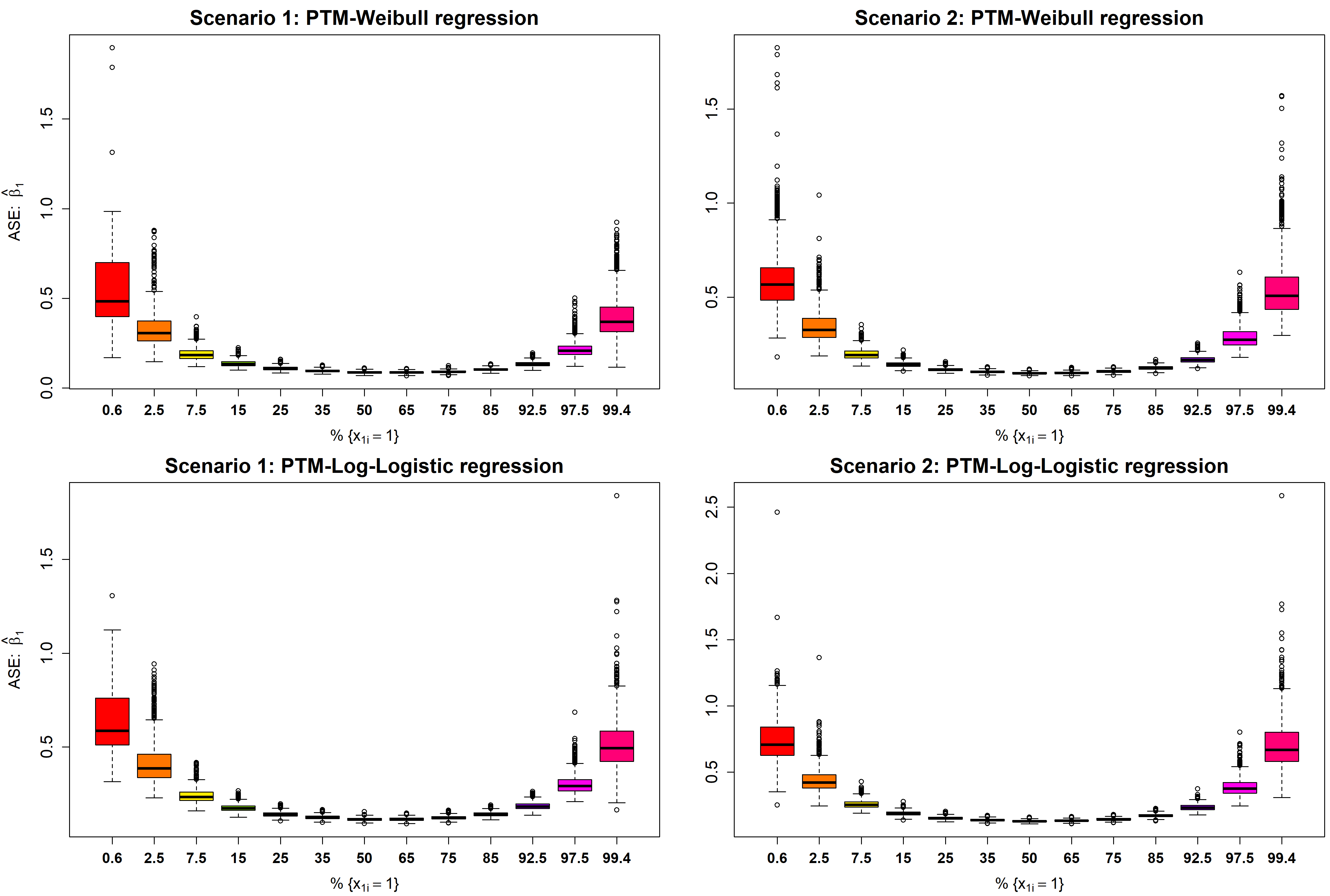}
\vspace{-0.2cm}
\caption{Main performance measures for the quantities of interest under varying proportions of successes in the binary covariate, $\%\{x_{1i} = 1\}$, with $n = 1000$ fixed.}
\label{perform1}
\end{figure}

\begin{figure}[H]
\centering
\includegraphics[scale=0.4]{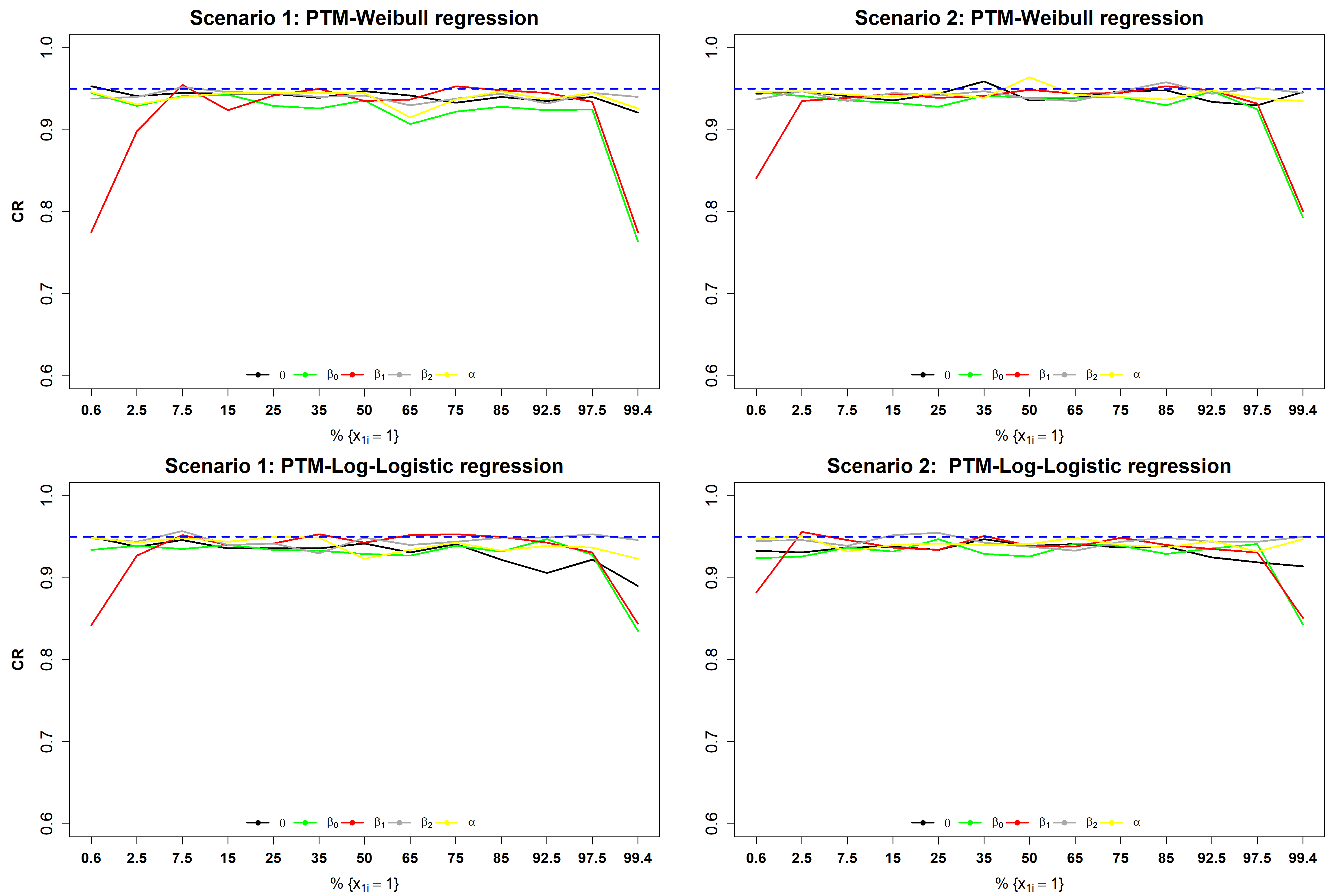}\\
\includegraphics[scale=0.4]{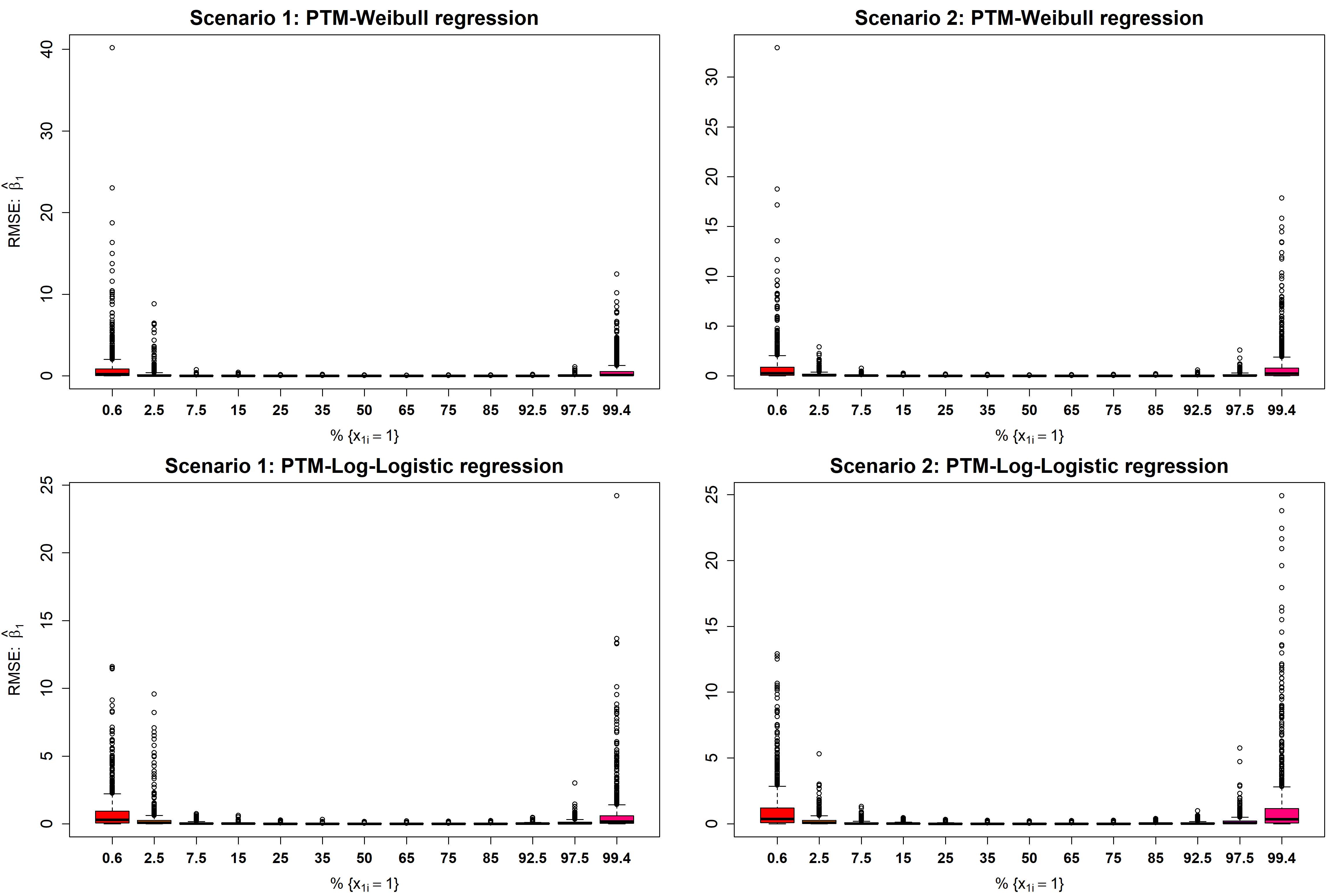}
\vspace{-0.2cm}
\caption{Main performance measures for the quantities of interest under varying proportions of successes in the binary covariate, $\%\{x_{1i} = 1\}$, with $n = 1000$ fixed.}
\label{perform2}
\end{figure}

Figure~\ref{fig:balanced} presents the proportion of samples affected by the ML problem across different configurations of the binary covariate. Enhancing the balance of this covariate not only significantly improved the model's overall performance metrics but also markedly reduced the incidence of the aforementioned phenomenon in the simulated samples. For instance, increasing the proportion of successes in $x_{1i}$ from 0.6\% to 2.5\% resulted in a substantial decline in the proportion of samples exhibiting ML issues. When the success rate was 0.6\%, ML occurred in 40\% and 20\% of samples under Scenarios 1 and 2, respectively. In contrast, increasing the success rate to 2.5\% sharply reduced these proportions to 2.7\% and 0\%. These findings further demonstrate that even modest improvements in the balance of the binary covariate can substantially mitigate the frequency of the ML problem.

\begin{figure}[H]
    \centering
    \includegraphics[scale=0.35]{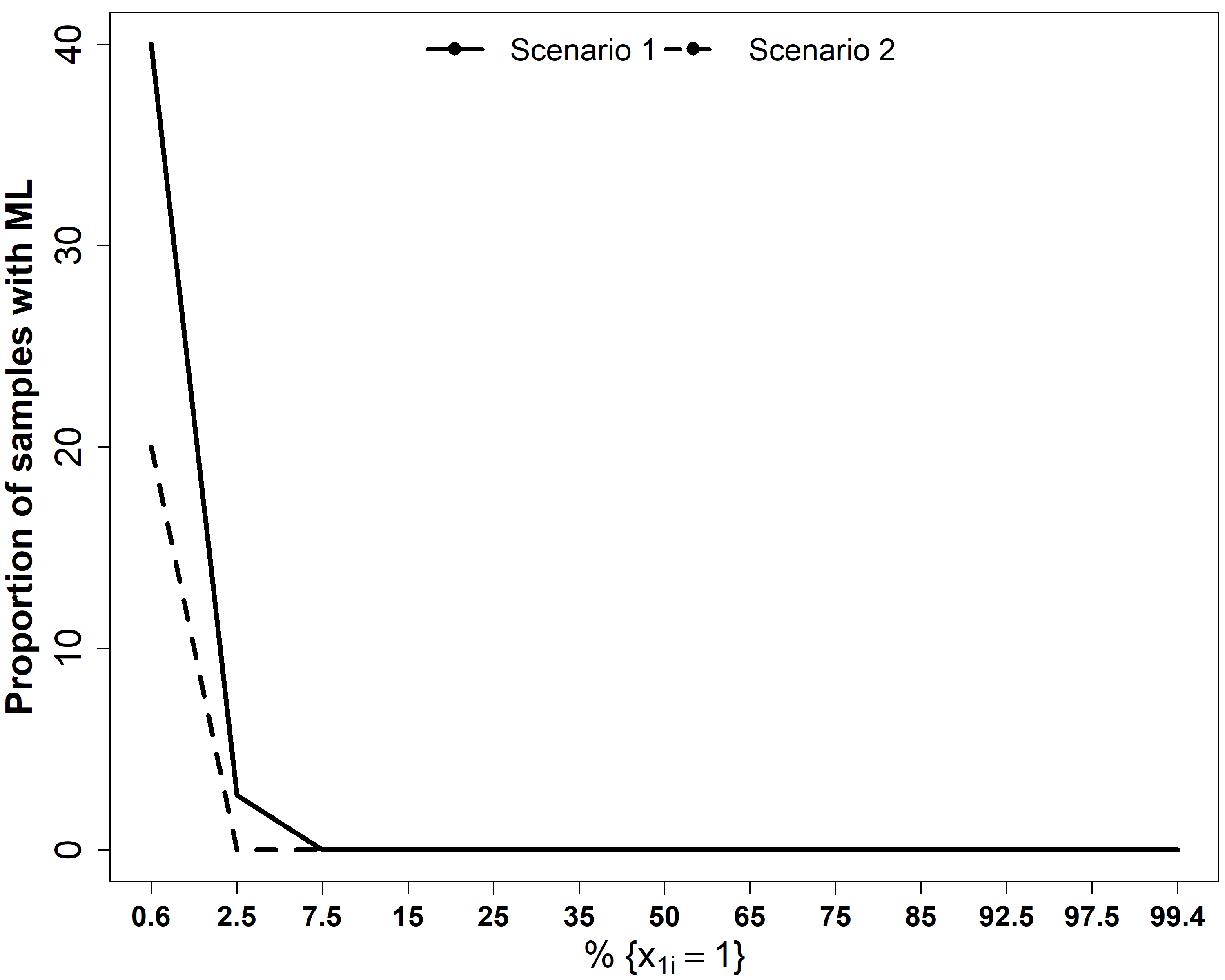}
    \vspace{-0.3cm}
    \caption{Behavior of the proportion of samples exhibiting ML under different balancing configurations of $x_{1i}$.}
    \label{fig:balanced}
\end{figure}

\section{Analysis of melanoma dataset}\label{sec4}

The melanoma dataset analyzed in this study consists of observations from patients diagnosed with primary melanoma and treated at the Hospital das Clínicas of the Federal University of Minas Gerais (HC-UFMG) and the Private Oncology Service for Digestive Tract Diseases of Belo Horizonte (ONCAD-BH) between 1995 and 2012. The study spanned 17 years. The time origin was defined as the date of melanoma diagnosis, and the end time as either the date of metastasis detection, indicating the event of interest, or the date of the last follow-up visit. Individuals who did not develop metastasis during follow-up were considered right-censored observations.
The dataset comprises sociodemographic and clinical information for approximately 514 patients. However, only 215 patients have complete data for all covariates, representing 41.83\% of the total; consequently, only these patients will be included in the analyses. 

The primary objective of this study was to examine the impact of various risk factors on disease progression. These risk factors were classified as either sociodemographic or clinical. Sociodemographic variables included sex (male and female) and patient age, with age stratified into three groups: 18--40 years, 41--60 years, and over 60 years. For the clinical factors (tumor characteristics), the following variables were evaluated: primary melanoma location (head, neck and trunk; upper and lower limbs; acral region); histological type (superficial spreading and lentigo maligna, nodular, and acral); ulceration (present or absent); and mitotic activity (present or absent).

A descriptive analysis of the melanoma dataset revealed that, among the total sample, 31 patients (14.42\%) developed metastasis, while 184 (85.58\%) were right-censored, indicating a high degree of censoring. The mean and median of patients who experienced metastasis are 8.75 and 4.60 weeks, respectively. The relative frequency distribution across categories was as follows: Gender (male: 37.21\%, female: 62.79\%); Age group (18--40 years: 26.05\%, 41--60 years: 40.47\%, over 60 years: 33.49\%); Tumor primary site (head, neck+trunk: 53.95\%, upper and lower limbs: 28.37\%, acral region: 17.67\%); Histological type (superficial spreading+lentigo maligna: 72.56\%, nodular: 16.74\%, acral: 10.70\%); Ulceration (present: 20.0\%, absent: 80.0\%); and Mitotic activity (present: 64.19\%, absent: 35.81\%). Notably, all patients with mitotic activity developed metastasis (the event under study). This finding demonstrates perfect separation for this covariate and further underscores the ML problem present in the melanoma dataset, as discussed in Section~\ref{sec1}.

\vspace{-0.2cm}
\begin{figure}[H]
	\centering
	\includegraphics[scale=0.45]{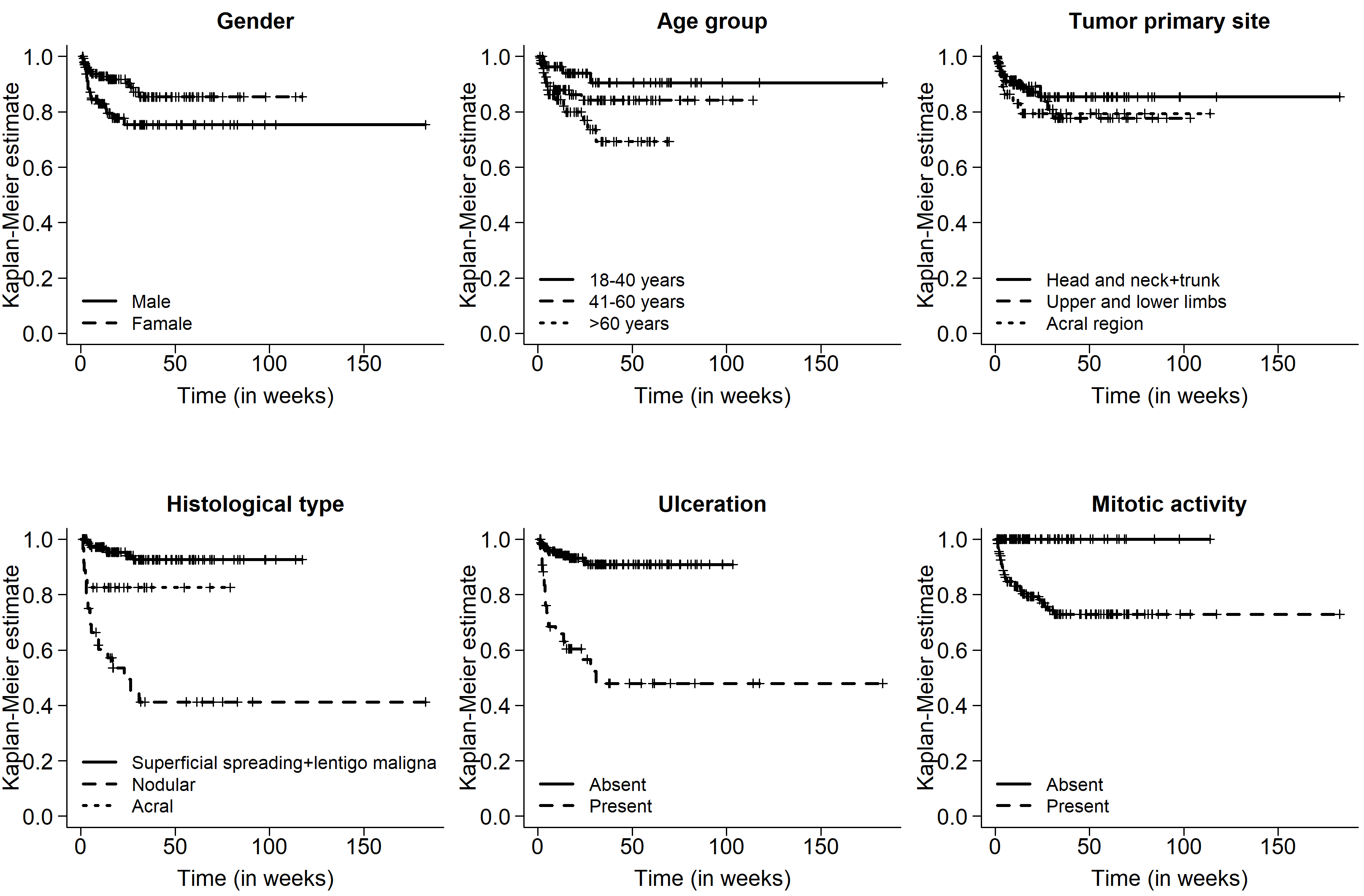}
    \vspace{-0.2cm}
	\caption{Kaplan--Meier curves for each risk factor considered in the analysis.}
	\label{fig:km_covariaveis}
\end{figure}

The survival curves in Figure~\ref{fig:km_covariaveis} complement the descriptive analysis. There, the KM estimates are displayed according to patients’ sociodemographic and clinical characteristics. Notably, higher survival probabilities are seen among female patients, those aged 18--40 years, individuals with tumors located in the head, neck, and trunk, patients with superficial spreading or lentigo maligna histological types, and those without ulceration or mitotic activity. To assess the impact of the ML problem on model fitting and to ensure the interpretability of the resulting estimates, we compared results obtained using the standard (unmodified) score function with those derived from the modified score function incorporating Firth’s correction.

The results of the model fitting performed in this study are presented in Table~\ref{tab:APL}. In this table, the reference level “Supspread+lentMaligna” for the histological type variable corresponds to the superficial spreading + lentigo maligna category. Notably, the magnitude of the standard errors for $\hat{\beta}_{\text{mt}}$ underscores the extent to which the ML problem can compromise the estimation of the coefficient associated with mitotic activity, which is directly implicated in the issue described above. The low precision of the coefficient estimate compels the Wald test statistic, $z = \hat{\beta}_{\text{mt}}/SE(\hat{\beta}_{\text{mt}})$, to approach zero, yielding $p$-values close to 1 and, consequently, a failure to reject the null hypothesis of no covariate effect. Moreover, the instability observed in estimating the coefficient for mitotic activity extends to the intercept estimates in both models. This phenomenon may be attributed to the fact that, as these intercepts are not linked to any specific covariate, they tend to capture the overall instability of the estimation procedure, as evidenced by their large standard errors.

\begin{table}[H]
	\centering
    \captionsetup{
    font={sf},
    justification=raggedright,
    singlelinecheck=false}
	\caption{Standard and modified MLEs, standard errors (SEs), and $p$-values for PTM-Weibull and Log-Logistic regression.}
   \vspace{-0.2cm}
	\label{tab:APL}
	\sf
    \small
	\renewcommand{\arraystretch}{1.08}
	\setlength{\tabcolsep}{0.10cm}
	\begin{tabular}{llrrrrrr}
	\toprule
	\multicolumn{8}{c}{\normalsize \sf PTM-Weibull regression}\\
	\toprule
	\multirow{2}{*}{Risk factor}& \multirow{2}{*}{Category} & \multicolumn{3}{c}{\sf Standard score function}
	& \multicolumn{3}{c}{\bf \sf Modified score function} \\
	\cmidrule(lr){3-5}\cmidrule(lr){6-8}
	&  & Estimate & SE & $p$-value & Estimate & SE & $p$-value\\
	\midrule
	Intercept & & 22.694 & 506.322 & 0.964 & 5.967 & 0.784 & $<$0.001 \\
			
 Gender & Male & &  & & & & \\
		& Female& 1.714 & 0.591 & 0.004 & 0.406 & 0.395 & 0.304 \\
			
 Age group& 18--40 years & &  & & & & \\
		   & 41--60 years &-0.304 & 0.859 & 0.723 & 0.283 & 0.480 & 0.555\\
		   & $>$60 years  &-0.903 & 0.876 & 0.303& 0.517 & 0.569 & 0.364 \\
			
 Tumor primary site
		& Head and neck+trunk   & &  & & & & \\
		& Upper and lower limbs &-0.902 & 0.715 & 0.207&-0.219 & 0.407 & 0.591 \\
		& Acral region &-0.219 & 0.731 & 0.764 &-0.439 & 0.517 & 0.396 \\
			
Histological type & Supspread+lentMaligna& &  & & & & \\
			& Nodular &-2.459 & 0.696 & $<$0.001 & $-2.371$ & 0.475 & $<$0.001 \\
			& Acral&-1.001& 0.959 & 0.296&-3.701 & 0.660 & $<$0.001 \\
			
Ulceration
            & Absent  & &  & & & & \\
			& Present &-1.691 & 0.594 & 0.004 &-0.420 & 0.392 & 0.285 \\
			
Mitotic activity
			& Absent & &  & & & & \\
			& Present&-16.018 & 506.318 & 0.975& -1.407 & 0.752 & 0.061 \\
			$\theta$ & & 1.355 & 0.426 & 0.001 & 0.590 & 0.138 & $<$0.001 \\
			$\alpha$ & & 0.873 & 0.142 & $<$0.001 & 1.300 & 0.161 & $<$0.001 \\
			\toprule
\multicolumn{8}{c}{\normalsize\sf PTM-Log-Logistic regression}\\
\toprule
\multirow{2}{*}{Risk factor}& \multirow{2}{*}{Category} & \multicolumn{3}{c}{\sf Standard score function}	& \multicolumn{3}{c}{\bf \sf Modified score function} \\
\cmidrule(lr){3-5}\cmidrule(lr){6-8}
	&  & Estimate & SE & $p$-value & Estimate & SE & $p$-value\\
	\midrule
	Intercept & & 20.249 & 365.083 & 0.956 & 7.578 & 1.302 & $<$0.001 \\
	Gender  
            & Male & &  & & & & \\
			& Female & 1.462 & 0.587 & 0.013 & 1.008 & 0.465 & 0.030 \\
	Age group
			  & 18--40 years & &  & & & & \\
			& 41--60 years & -0.559 & 0.805 & 0.487 & -0.417 & 0.560 & 0.457 \\
			& $>$60 years  & -0.941 & 0.805 & 0.242 & -0.473 & 0.527 & 0.370 \\
			
Tumor primary site
		& Head and neck+trunk & &  & & & & \\
		& Upper and lower limbs &-1.069 & 0.695 & 0.124 &-0.784 & 0.496 & 0.114 \\
		& Acral region &-0.053 & 0.717 & 0.941 &-0.133 & 0.529 & 0.802 \\
			
Histological type
	    & Supspread+lentMaligna & &  & & & & \\
		& Nodular & -2.343 & 0.646 & $<$0.001 & -1.919 & 0.479 & $<$0.001\\
		& Acral   &-1.094 & 0.927 & 0.238 &-1.633 & 0.750 & 0.029 \\
			
Ulceration
	    & Absent & &  & & & & \\
		& Present &-1.804 & 0.572 & 0.002 & -1.333 & 0.435 & 0.002 \\
		
Mitotic activity
		& Absent  & &  & & & & \\
		& Present &-13.673 & 365.076 & 0.970 &-2.285 & 1.134 & 0.044 \\
        
        $\theta$ & & 1.415 & 0.535 & 0.008 & 1.152 & 0.429 & 0.007 \\
		$\alpha$ & & 1.001 & 0.205 & $<$0.001 & 1.379 & 0.249 & $<$0.001\\
		\bottomrule
		\end{tabular}
\end{table}

It is important to emphasize that when the standard (unmodified) score function is employed, the reported estimates correspond to the values produced by the algorithm upon reaching the maximum number of iterations, as the iterative procedure typically fails to satisfy the convergence criterion in such cases. Consequently, the log-likelihood specified in Equation~\ref{ObsLikelihood} is not maximized at $\hat{\beta}_{\text{mt}}$.

\begin{table}[H]
\raggedleft
\begin{threeparttable}
\captionsetup{
 font={sf},
 justification=raggedright,
 singlelinecheck=false}
\caption{Information criterion values for the fitted models. Here, AIC and BIC denote the Akaike and Bayesian information criteria, respectively.}
\vspace{-0.2cm}
\label{InfoCriteria}
\setlength{\tabcolsep}{0.2cm}
{\sf
\begin{tabular}{lcc}
\toprule
Regression model & AIC & BIC \\
\midrule
PTM-Weibull      & 316.829 & 357.276 \\
PTM-Log-Logistic & 308.224 & 348.672 \\
\bottomrule
\end{tabular}
}
\end{threeparttable}
\end{table}

The model fitting results indicate that modifying the score function enabled finite and more precise estimates for both the intercept term and the coefficient associated with the mitosis factor. More specifically, in the PTM-Weibull regression, the modified MLE of the coefficient for the mitosis factor, denoted as $\beta_{mt}$, and its corresponding standard error changed from $\hat{\beta}_{mt} = -16.018$ ($\mbox{SE}_{\hat{\beta}_{\text{mt}}}$ = 506.318) when using the standard score function to $\hat{\beta}_{mt}^{*} = -1.407$ ($\mbox{SE}_{\hat{\beta}_{\text{mt}}}$ = 0.752) with the modified score function. Similarly, in the PTM-Log-Logistic regression, the estimates changed from $\hat{\beta}_{mt} = -13.673$ ($\mbox{SE}_{\hat{\beta}_{\text{mt}}}$ = 365.076) with the standard score function to $\hat{\beta}_{mt}^{*} = -2.285$ ($\mbox{SE}_{\hat{\beta}_{\text{mt}}}$ = 1.134) with the modified approach.

\begin{figure}[H]
\centering
\includegraphics[scale=0.55]{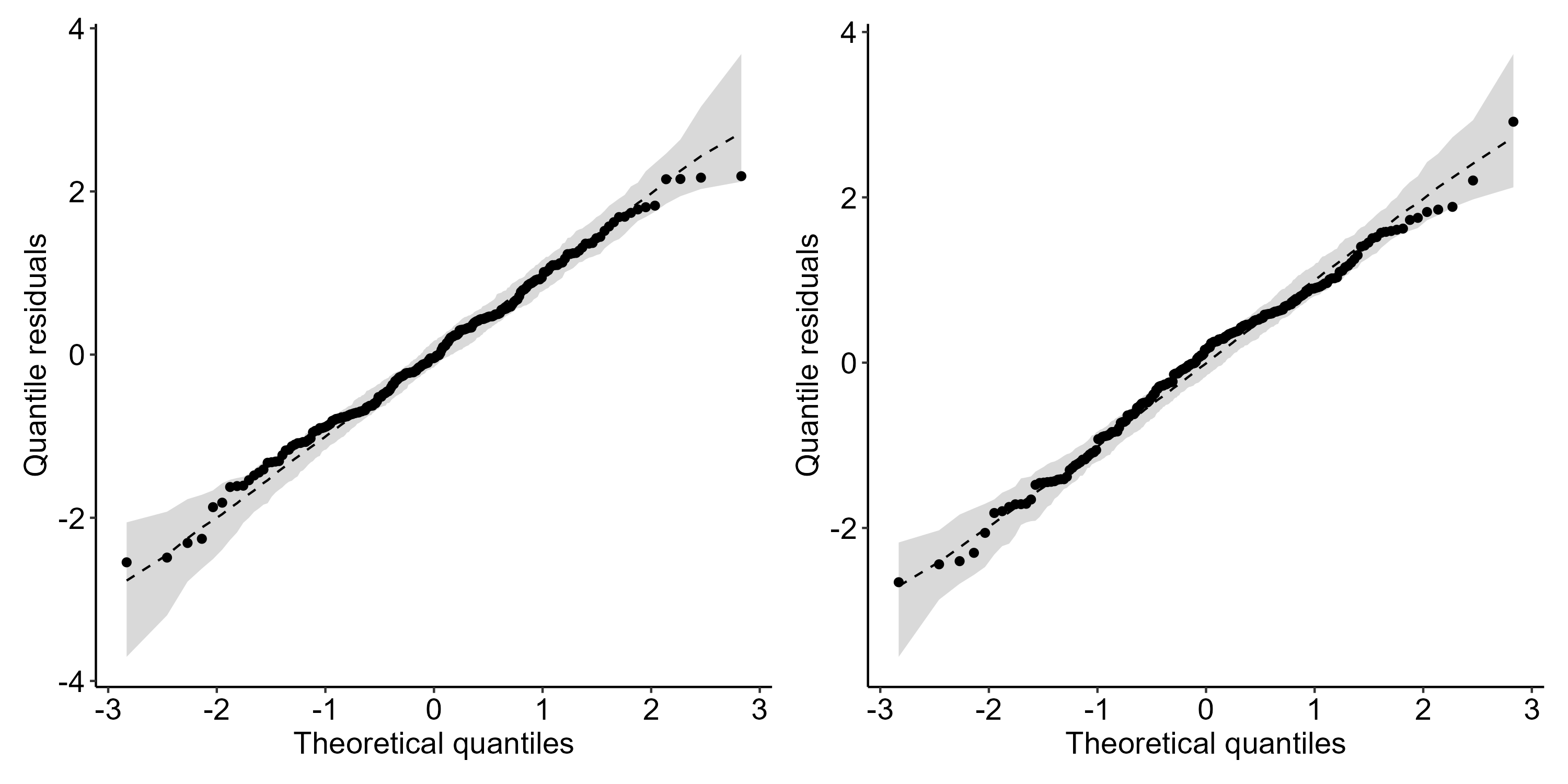}
\vspace{-0.3cm}
\caption{Quantile residual plots for the PTM-Log-Logistic models using standard (left) and modified (right) score functions.}
\label{fig:residuals}
\end{figure}

In both models, the mitotic activity factor was not statistically significant when the standard score function was used. The p-values were close to 1, and the corresponding confidence intervals were extremely wide, reflecting substantial imprecision in the estimates. However, after modifying the score function, this factor became statistically significant in the PTM-Log-Logistic regression ($p$-value = 0.044). Despite the improvements observed in the parameter estimates, the mitosis risk factor remained non-significant in the PTM-Weibull regression ($p$-value = 0.061) at the 5\% significance level. Notably, a key finding from the comparison of parameter estimates across the two models and estimation procedures is the remarkable consistency observed for covariates not directly associated with the ML problem, regardless of whether the standard or modified score function was used.

Model fit was assessed using the AIC and BIC criteria, both calculated from the penalized log-likelihood functions. The corresponding values are presented in Table~\ref{InfoCriteria}. As lower AIC and BIC values indicate improved model fit, these results clearly favor the PTM-Log-Logistic regression for the melanoma dataset. Additionally, the quantile residuals in Figure~\ref{fig:residuals} show similar patterns under the standard and penalized models, with most observations closely following the reference line and falling within the confidence bands. Overall, penalization improves estimation stability without materially affecting model fit. Although minor deviations are observed in the tails, particularly in the upper tail under the modified score function, these departures are not substantial. Overall, the results indicate that incorporating penalization does not materially alter the distributional behavior of the residuals. These findings indicate that the modified estimation procedure enhances numerical stability without diminishing the overall adequacy of model fit.

Overall, the results of this section are broadly consistent with those reported in the oncology literature \citep{retsas2002prognostic,cherobin2018,chiu2026clinical}, both in terms of the statistical significance of the identified risk factors and the direction of their effects on time to metastasis in patients with primary cutaneous melanoma.

\section{Concluding remarks}\label{sec5}

This study introduces a novel modification to the score function in PTMs to address the persistent issue of nonexistence of finite parameter estimates due to ML, a challenge that, to the best of our knowledge, no studies have addressed this issue for this class of models. Specifically, the proposed methodology is developed for the PTM-Weibull and PTM-Log-Logistic regression, offering a more robust and reliable inference framework. 

Simulation studies indicated that the proposed methodology yields finite estimates for the coefficient associated with the binary covariate. However, severe imbalance in this covariate can compromise estimation precision, leading to greater relative bias and lower coverage rates. Furthermore, the balancing experiment confirmed that improved covariate balance contributes to substantial gains in performance measures, underscoring its critical importance for estimation accuracy. Overall, balanced configurations of the covariate $x_{1i}$, which was used to induce the ML problem, resulted in satisfactory estimation performance. In contrast, highly unbalanced configurations, characterized by a predominance of either successes or failures, resulted in poorer performance metrics, even when the phenomenon under investigation was absent. The findings also indicated that the estimation of parameters unrelated to the covariate inducing the ML problem generally remained robust. 

Application of the proposed methodology to the melanoma dataset yielded several notable findings. In particular, mitotic activity, a principal prognostic marker of tumor progression, demonstrated a statistically significant association. Furthermore, residual analysis confirmed the adequacy of the model fit. Collectively, these findings corroborate previous evidence reported in the oncology literature.

\section*{Acknowledgment}
The first author gratefully acknowledges the support of the Centro de Gestão e Estudos Estratégicos (CGEE). The second author gratefully acknowledges the support of the Fundação de Amparo à Pesquisa do Distrito Federal (FAPDF), Brazil, and the Decanato de Pós-Graduação (DPG) of the University of Brasília.

\section*{Conflict of interest}
The authors declare that they have no conflict of interest.

\section*{Data availability statement}
The data that support the findings of this study are available from the corresponding author upon reasonable request.

\bibliographystyle{apalike}
\bibliography{refs}

\appendix
\section*{Appendices}

\renewcommand{\thesection}{A\arabic{section}}
\renewcommand{\theequation}{A\arabic{equation}} 
\setcounter{equation}{0}
\renewcommand{\thefigure}{\thesection.\arabic{figure}}
\renewcommand{\thetable}{\thesection.\arabic{table}}
\counterwithin{figure}{section}\setcounter{figure}{0}
\counterwithin{table}{section}\setcounter{table}{0}

\section{Analytical expressions for the main cumulants}
\label{apendiceA1}
In this section of the appendix, we present the first-, second-, and third-order cumulants employed in the estimation procedures for the models considered in this paper. Without loss of generality, let $\ell\left(\bm{\psi}\right) = \ell\left(\bm{\psi} \mid D_o\right)$ denote the observed log-likelihood function, defined in terms of the Equation~\eqref{ObsLikelihood}, and let $\mathbf{x}_i=\left(x_{i0}, x_{i1}, \ldots, x_{iq_1}\right)^\top$ denote the covariate vector, where $x_{i0} = 1$ for all $i \in {1, 2, \ldots, n}$. Additionally, define $\bm{\beta} = \left(\beta_0, \beta_1, \ldots, \beta_{q_1}\right)^\top$ as the $q_1$-dimensional vector of regression coefficients. Below, we present the first-, second-, and third-order cumulants, each defined for all $r, s, k \in \{0, 1, 2, \ldots, q_1\}$, for the Weibull and Log-Logistic PTM regression.

\subsection{PTM-Weibull regression}
\renewcommand{\thetable}{\thesection.\arabic{table}}

\subsubsection{First-order cumulants}

\begin{eqnarray*}
	\frac{\partial \ell\left(\bm{\psi}\right)}{\partial \beta_r}&=&
	\sum_{i=1}^n
	\alpha x_{ir}
	\Bigg[\delta_i\left(
	\left(\frac{t_i}{\exp(\mathbf{x}_i ^ \top \boldsymbol{\beta})}\right)^\alpha-1
	\right) +\theta\,
	\exp\!\left[-\left(\frac{t_i}{\exp(\mathbf{x}_i ^ \top \boldsymbol{\beta})}\right)^\alpha\right]
	\left(\frac{t_i}{\exp(\mathbf{x}_i ^ \top \boldsymbol{\beta})}\right)^\alpha
	\Bigg],
	\\
	\frac{\partial \ell\left(\bm{\psi}\right)}{\partial \theta}
	&=&\sum_{i=1}^n\left[
	\frac{\delta_i}{\theta}
	-
	\left(1-
	\exp\!\left[-\left(\frac{t_i}{\exp(\mathbf{x}_i ^ \top \boldsymbol{\beta})}\right)^\alpha\right]
	\right)\right],
	\\
	\frac{\partial \ell\left(\bm{\psi}\right)}{\partial \alpha}&=&\sum_{i=1}^n
	\Bigg[\delta_i\Bigg(
	\frac{1}{\alpha}
	+\log\left(\frac{t_i}{\exp(\mathbf{x}_i ^ \top \boldsymbol{\beta})}\right)
	-\left(\frac{t_i}{\exp(\mathbf{x}_i ^ \top \boldsymbol{\beta})}\right)^\alpha
	\log\left(\frac{t_i}{\exp(\mathbf{x}_i ^ \top \boldsymbol{\beta})}\right)
	\Bigg) \\
	&&-\theta\, \exp\!\left[-\left(\frac{t_i}{\exp(\mathbf{x}_i ^ \top \boldsymbol{\beta})}\right)^\alpha\right]
	\left(\frac{t_i}{\exp(\mathbf{x}_i ^ \top \boldsymbol{\beta})}\right)^\alpha
	\log\left(\frac{t_i}{\exp(\mathbf{x}_i ^ \top \boldsymbol{\beta})}\right)
	\Bigg].
\end{eqnarray*}

\subsubsection{Second-order cumulants}

	\begin{eqnarray*}
		\frac{\partial^2\ell\left(\bm{\psi}\right)}{\partial\beta_r\,\partial\beta_s}
		&=& - \sum_{i=1}^n \alpha^2 x_{ir}x_{is} \left(\frac{t_i}{\exp(\mathbf{x}_i ^ \top \boldsymbol{\beta})}\right)^\alpha
		\Bigg[\delta_i + \theta\, \exp\!\left[-\left(\frac{t_i}{\exp(\mathbf{x}_i ^ \top \boldsymbol{\beta})}\right)^\alpha\right] \left(1-\left(\frac{t_i}{\exp(\mathbf{x}_i ^ \top \boldsymbol{\beta})}\right)^\alpha\right) \Bigg],
		\\
		\frac{\partial^2\ell\left(\bm{\psi}\right)}{\partial\theta^2} &=& -\sum_{i=1}^n\frac{\delta_i}{\theta^2},
		\\
		\frac{\partial^2\ell\left(\bm{\psi}\right)}{\partial\alpha^2} &=& \sum_{i=1}^n \Bigg[-\frac{\delta_i}{\alpha^2} - \left(\log\left(\frac{t_i}{\exp(\mathbf{x}_i ^ \top \boldsymbol{\beta})}\right)\right)^2 \left(\frac{t_i}{\exp(\mathbf{x}_i ^ \top \boldsymbol{\beta})}\right)^\alpha \times \\
		&& \Bigg\{ \delta_i + \theta\, \exp\!\left[-\left(\frac{t_i}{\exp(\mathbf{x}_i ^ \top \boldsymbol{\beta})}\right)^\alpha\right] \left(1-\left(\frac{t_i}{\exp(\mathbf{x}_i ^ \top \boldsymbol{\beta})}\right)^\alpha\right) \Bigg\} \Bigg].
	\end{eqnarray*}

	\begin{eqnarray*}
		\frac{\partial^2\ell\left(\bm{\psi}\right)}{\partial\theta\,\partial\beta_r}
		&=& \sum_{i=1}^n \alpha x_{ir}\, \exp\!\left[-\left(\frac{t_i}{\exp(\mathbf{x}_i ^ \top \boldsymbol{\beta})}\right)^\alpha\right] \left(\frac{t_i}{\exp(\mathbf{x}_i ^ \top \boldsymbol{\beta})}\right)^\alpha,
		\\
		\frac{\partial^2\ell\left(\bm{\psi}\right)}{\partial\theta\,\partial\alpha}
		&=& - \sum_{i=1}^n \left(\frac{t_i}{\exp(\mathbf{x}_i ^ \top \boldsymbol{\beta})}\right)^\alpha \log\left(\frac{t_i}{\exp(\mathbf{x}_i ^ \top \boldsymbol{\beta})}\right) \exp\!\left[-\left(\frac{t_i}{\exp(\mathbf{x}_i ^ \top \boldsymbol{\beta})}\right)^\alpha\right],
		\\
		\frac{\partial^2\ell\left(\bm{\psi}\right)}{\partial\beta_r\,\partial\alpha}
		&=& \sum_{i=1}^n x_{ir} \Bigg[\delta_i\left( \left(\frac{t_i}{\exp(\mathbf{x}_i ^ \top \boldsymbol{\beta})}\right)^\alpha-1 \right) + \theta\, \exp\!\left[-\left(\frac{t_i}{\exp(\mathbf{x}_i ^ \top \boldsymbol{\beta})}\right)^\alpha\right] \left(\frac{t_i}{\exp(\mathbf{x}_i ^ \top \boldsymbol{\beta})}\right)^\alpha \\
		&& +\alpha \log\left(\frac{t_i}{\exp(\mathbf{x}_i ^ \top \boldsymbol{\beta})}\right) \left(\frac{t_i}{\exp(\mathbf{x}_i ^ \top \boldsymbol{\beta})}\right)^\alpha \Bigg\{ \delta_i + \theta\, \exp\!\left[-\left(\frac{t_i}{\exp(\mathbf{x}_i ^ \top \boldsymbol{\beta})}\right)^\alpha\right]\left( 1-\left(\frac{t_i}{\exp(\mathbf{x}_i ^ \top \boldsymbol{\beta})}\right)^\alpha \right) \Bigg\} \Bigg].
	\end{eqnarray*}

\noindent Therefore, the observed information matrix (variance-covariance matrix) is given by $\mathcal{H}\left(\bm{\psi}\right)=\left(-\frac{\partial^2 \ell\left(\bm{\psi}\right)}{\partial \bm{\psi}^\top \partial\bm{\psi}}\right)_{d\times d}$, where $u, b \in \{1, 2, \ldots, d\}$, and $d=q_{1}+3$ denotes the dimension of the parameter vector $\bm{\psi}$.

\subsubsection{Third-order cumulants}

The third-order cumulants are obtained by differentiating the $(u, b)$-th entry of $\mathcal{H}(\bm{\psi})$ with respect to $\psi_\nu$, for all $\nu \in \{1, 2, \ldots, d\}$. That is, $
A_\nu(\bm{\psi}) = \frac{1}{2} \sum\limits_{u=1}^d \sum\limits_{b=1}^d \mathcal{H}^{u, b} \left( \frac{\partial \mathcal{H}{u, b}}{\partial \psi{\nu}} \right) = \frac{1}{2} \sum\limits_{u=1}^d \sum\limits_{b=1}^d \mathcal{H}^{u, b} \xi_{u, b, \nu}$, where $\mathcal{H}_{u, b}$ and $\mathcal{H}^{u, b}$ denote the $(u, b)$-th entries of $\mathcal{H}(\bm{\psi})$ and its inverse $\mathcal{H}^{-1}(\bm{\psi})$, respectively. The explicit expressions for the third-order cumulants, $\xi_{u, b, \nu}$, are presented below.

	\begin{eqnarray*}
	\frac{\partial^3\ell\left(\bm{\psi}\right)}{\partial\theta^3}&=&
		2\sum_{i=1}^n\frac{\delta_i}{\theta^3}, \quad \frac{\partial^3\ell}{\partial\theta^2\partial\beta_r}=0,
		\quad
		\frac{\partial^3\ell\left(\bm{\psi}\right)}{\partial\theta^2\partial\alpha}=0,
		\\
		\frac{\partial^3\ell\left(\bm{\psi}\right)}{\partial\theta\,\partial\beta_r\,\partial\beta_s}
		&=&
		-\sum_{i=1}^n
		\alpha^2 x_{ir}x_{is}
		\left(\frac{t_i}{\exp(\mathbf{x}_i^\top \boldsymbol{\beta})}\right)^\alpha
		\exp\!\left[
		-\left(\frac{t_i}{\exp(\mathbf{x}_i^\top \boldsymbol{\beta})}\right)^\alpha
		\right] \times
		\left[
		1-\left(\frac{t_i}{\exp(\mathbf{x}_i^\top \boldsymbol{\beta})}\right)^\alpha
		\right],
		\\
		\frac{\partial^3\ell\left(\bm{\psi}\right)}{\partial\theta\,\partial\beta_r\,\partial\alpha}
		&=&
		\sum_{i=1}^n
		x_{ir}
		\left(\frac{t_i}{\exp(\mathbf{x}_i^\top \boldsymbol{\beta})}\right)^\alpha
		\exp\!\left[
		-\left(\frac{t_i}{\exp(\mathbf{x}_i^\top \boldsymbol{\beta})}\right)^\alpha
		\right] \\
        &\times&
		\Bigg[1+\alpha
		\log\left(\frac{t_i}{\exp(\mathbf{x}_i^\top \boldsymbol{\beta})}\right)
		\left\{
		1-\left(\frac{t_i}{\exp(\mathbf{x}_i^\top \boldsymbol{\beta})}\right)^\alpha
		\right\}
		\Bigg],
		\\
		\frac{\partial^3\ell\left(\bm{\psi}\right)}{\partial\theta\,\partial\alpha^2}
		&=&
		-\sum_{i=1}^n
		\left(\frac{t_i}{\exp(\mathbf{x}_i^\top \boldsymbol{\beta})}\right)^\alpha
		\left[
		\log\left(\frac{t_i}{\exp(\mathbf{x}_i^\top \boldsymbol{\beta})}\right)
		\right]^2
		\exp\!\left[
		-\left(\frac{t_i}{\exp(\mathbf{x}_i^\top \boldsymbol{\beta})}\right)^\alpha
		\right]\times\left[1-\left(\frac{t_i}{\exp(\mathbf{x}_i^\top \boldsymbol{\beta})}\right)^\alpha\right],\\
    \frac{\partial^3\ell\left(\bm{\psi}\right)}{\partial\alpha^3}
		&=&
		\sum_{i=1}^n
		\Bigg[
		\frac{2\delta_i}{\alpha^3}
		-
		\left(\frac{t_i}{\exp(\mathbf{x}_i^\top \boldsymbol{\beta})}\right)^\alpha
		\left[
		\log\left(\frac{t_i}{\exp(\mathbf{x}_i^\top \boldsymbol{\beta})}\right)
		\right]^3 \times
		\\
		&&
		\Bigg\{
		\delta_i
		+
		\theta
		\exp\!\left[
		-\left(\frac{t_i}{\exp(\mathbf{x}_i^\top \boldsymbol{\beta})}\right)^\alpha
		\right]
		\left[
		1
		-
		3\left(\frac{t_i}{\exp(\mathbf{x}_i^\top \boldsymbol{\beta})}\right)^\alpha
		+
		\left(\frac{t_i}{\exp(\mathbf{x}_i^\top \boldsymbol{\beta})}\right)^{2\alpha}\right]\Bigg\}\Bigg],
	\end{eqnarray*}

	\begin{eqnarray*}
		\frac{\partial^3\ell\left(\bm{\psi}\right)}{\partial\beta_r\,\partial\beta_s\,\partial\beta_k}
		&=&
		\sum_{i=1}^n
		\alpha^3 x_{ir}x_{is}x_{ik}
		\left(\frac{t_i}{\exp(\mathbf{x}_i^\top \boldsymbol{\beta})}\right)^\alpha \times
		\\
		&& 
		\Bigg[
		\delta_i
		+
		\theta
		\exp\!\left[
		-\left(\frac{t_i}{\exp(\mathbf{x}_i^\top \boldsymbol{\beta})}\right)^\alpha
		\right]
		\left\{
		1
		-
		3\left(\frac{t_i}{\exp(\mathbf{x}_i^\top \boldsymbol{\beta})}\right)^\alpha
		+
		\left(\frac{t_i}{\exp(\mathbf{x}_i^\top \boldsymbol{\beta})}\right)^{2\alpha}
		\right\}
		\Bigg],
		\\
		\frac{\partial^3\ell\left(\bm{\psi}\right)}{\partial\beta_r\,\partial\beta_s\,\partial\alpha}
		&=&
		-\sum_{i=1}^n
		\alpha x_{ir}x_{is}
		\left(\frac{t_i}{\exp(\mathbf{x}_i^\top \boldsymbol{\beta})}\right)^\alpha 
		\\
		&&  \times
		\Bigg[
		\delta_i
		\left\{
		2
		+
		\alpha
		\log\left(\frac{t_i}{\exp(\mathbf{x}_i^\top \boldsymbol{\beta})}\right)
		\right\}
		\\
		&& +
		\theta
		\exp\!\left[
		-\left(\frac{t_i}{\exp(\mathbf{x}_i^\top \boldsymbol{\beta})}\right)^\alpha
		\right]
		\Bigg\{
		2\left[
		1-\left(\frac{t_i}{\exp(\mathbf{x}_i^\top \boldsymbol{\beta})}\right)^\alpha
		\right]
		\\
		&& +
		\alpha
		\log\left(\frac{t_i}{\exp(\mathbf{x}_i^\top \boldsymbol{\beta})}\right)
		\left[
		1
		-
		3\left(\frac{t_i}{\exp(\mathbf{x}_i^\top \boldsymbol{\beta})}\right)^\alpha
		+
		\left(\frac{t_i}{\exp(\mathbf{x}_i^\top \boldsymbol{\beta})}\right)^{2\alpha}
		\right]
		\Bigg\}
		\Bigg],
		\\
		\frac{\partial^3\ell\left(\bm{\psi}\right)}{\partial\beta_r\,\partial\alpha^2}
		&=&
		\sum_{i=1}^n
		x_{ir}
		\left(\frac{t_i}{\exp(\mathbf{x}_i^\top \boldsymbol{\beta})}\right)^\alpha
		\log\left(\frac{t_i}{\exp(\mathbf{x}_i^\top \boldsymbol{\beta})}\right)
		\\
		&&\quad \times
		\Bigg[\delta_i\left\{2+\alpha
		\log\left(\frac{t_i}{\exp(\mathbf{x}_i^\top \boldsymbol{\beta})}\right)
		\right\}
		\\
		&&\quad +
		\theta
		\exp\!\left[
		-\left(\frac{t_i}{\exp(\mathbf{x}_i^\top \boldsymbol{\beta})}\right)^\alpha
		\right]
		\Bigg\{
		2\left[
		1-\left(\frac{t_i}{\exp(\mathbf{x}_i^\top \boldsymbol{\beta})}\right)^\alpha
		\right]\\
		&&\quad +
		\alpha
		\log\left(\frac{t_i}{\exp(\mathbf{x}_i^\top \boldsymbol{\beta})}\right)
		\left[1-3\left(\frac{t_i}{\exp(\mathbf{x}_i^\top \boldsymbol{\beta})}\right)^\alpha+
		\left(\frac{t_i}{\exp(\mathbf{x}_i^\top \boldsymbol{\beta})}\right)^{2\alpha}
		\right]\Bigg\}\Bigg].
	\end{eqnarray*}

\vspace{0.5cm}
\subsection{PTM-Log-Logistic regression}
\renewcommand{\thetable}{\thesection.\arabic{table}}

\subsubsection{First-order cumulants}
	\begin{eqnarray*}
			\frac{\partial \ell\left(\bm{\psi}\right)}{\partial \theta}&=&
			\sum_{i=1}^n\left(\frac{\delta_i}{\theta}-\frac{\left(\frac{t_i}{\exp(\mathbf{x}_i ^ \top \boldsymbol{\beta})}\right)^\alpha}{1+\left(\frac{t_i}{\exp(\mathbf{x}_i ^ \top \boldsymbol{\beta})}\right)^\alpha}
			\right),
			\\
			\frac{\partial \ell\left(\bm{\psi}\right)}{\partial \beta_r}&=&\sum_{i=1}^n x_{ir}
			\left[-\delta_i\alpha + \frac{2\delta_i\alpha
				\left(\frac{t_i}{\exp(\mathbf{x}_i ^ \top \boldsymbol{\beta})}\right)^\alpha}
			{1+\left(\frac{t_i}{\exp(\mathbf{x}_i ^ \top \boldsymbol{\beta})}\right)^\alpha} + \theta
			\frac{\alpha \left(\frac{t_i}{\exp(\mathbf{x}_i ^ \top \boldsymbol{\beta})}\right)^\alpha}
			{\left(1+\left(\frac{t_i}{\exp(\mathbf{x}_i ^ \top \boldsymbol{\beta})}\right)^\alpha\right)^2}
			\right],
			\\
			\frac{\partial \ell\left(\bm{\psi}\right)}{\partial \alpha}&=&\sum_{i=1}^n
			\Bigg[\delta_i \left( \frac{1}{\alpha} + \log t_i-\mathbf{x}_i ^ \top \boldsymbol{\beta}-\frac{
				2\left(\frac{t_i}{\exp(\mathbf{x}_i ^ \top \boldsymbol{\beta})}\right)^\alpha
				\log\left(\frac{t_i}{\exp(\mathbf{x}_i ^ \top \boldsymbol{\beta})}\right)}{1+\left(\frac{t_i}{\exp(\mathbf{x}_i ^ \top \boldsymbol{\beta})}\right)^\alpha}-\theta\frac{\left(\frac{t_i}{\exp(\mathbf{x}_i ^ \top \boldsymbol{\beta})}\right)^\alpha
				\log\left(\frac{t_i}{\exp(\mathbf{x}_i ^ \top \boldsymbol{\beta})}\right)}{\left(1+\left(\frac{t_i}{\exp(\mathbf{x}_i ^ \top \boldsymbol{\beta})}\right)^\alpha\right)^2}\right)\Bigg].
		\end{eqnarray*}

\subsubsection{Second-order cumulants}

    \begin{eqnarray*}
    \frac{\partial^2\ell\left(\bm{\psi}\right)}{\partial\theta^2}&=&
		-\sum_{i=1}^n\frac{\delta_i}{\theta^2}
        \\
		\frac{\partial^2\ell\left(\bm{\psi}\right)}{\partial\theta\,\partial\beta_r}&=&
		\sum_{i=1}^nx_{ir}\alpha
			\left(\frac{t_i}{\exp(\mathbf{x}_i ^ \top \boldsymbol{\beta})}\right)^\alpha\left[1+\left(\frac{t_i}{\exp(\mathbf{x}_i ^ \top \boldsymbol{\beta})}\right)^\alpha\right]^{-2},
		\\
		\frac{\partial^2\ell\left(\bm{\psi}\right)}{\partial\theta\,\partial\alpha}&=&-\sum_{i=1}^n\left(\frac{t_i}{\exp(\mathbf{x}_i ^ \top \boldsymbol{\beta})}\right)^\alpha
			\log\left(\frac{t_i}{\exp(\mathbf{x}_i ^ \top \boldsymbol{\beta})}\right)\left[1+\left(\frac{t_i}{\exp(\mathbf{x}_i ^ \top \boldsymbol{\beta})}\right)^\alpha\right]^{-2},
		\\
		\frac{\partial^2\ell\left(\bm{\psi}\right)}{\partial\beta_r\,\partial\beta_s}&=&-\sum_{i=1}^n
		x_{ir}x_{is}\Bigg[
		\frac{2\delta_i\alpha^2
			\left(\frac{t_i}{\exp(\mathbf{x}_i ^ \top \boldsymbol{\beta})}\right)^\alpha}
		{\left(1+\left(\frac{t_i}{\exp(\mathbf{x}_i ^ \top \boldsymbol{\beta})}\right)^\alpha\right)^2}\Bigg]-\theta\sum_{i=1}^nx_{ir}x_{is}\Bigg[
		\frac{\alpha^2 \left(\frac{t_i}{\exp(\mathbf{x}_i ^ \top \boldsymbol{\beta})}\right)^\alpha
			\left(1-\left(\frac{t_i}{\exp(\mathbf{x}_i ^ \top \boldsymbol{\beta})}\right)^\alpha
			\right)}{\left(1+\left(\frac{t_i}{\exp(\mathbf{x}_i ^ \top \boldsymbol{\beta})}\right)^\alpha\right)^3}
		\Bigg],
		\\
		\frac{\partial^2\ell\left(\bm{\psi}\right)}{\partial\beta_r\,\partial\alpha}&=&\sum_{i=1}^n x_{ir}
		\Bigg[-\delta_i + \frac{2\delta_i \left(\frac{t_i}{\exp(\mathbf{x}_i ^ \top \boldsymbol{\beta})}\right)^\alpha}
		{1+\left(\frac{t_i}{\exp(\mathbf{x}_i ^ \top \boldsymbol{\beta})}\right)^\alpha}+\frac{
			2\delta_i\alpha \left(\frac{t_i}{\exp(\mathbf{x}_i ^ \top \boldsymbol{\beta})}\right)^\alpha
			\log\left(\frac{t_i}{\exp(\mathbf{x}_i ^ \top \boldsymbol{\beta})}\right)}{\left(1+\left(\frac{t_i}{\exp(\mathbf{x}_i ^ \top \boldsymbol{\beta})}\right)^\alpha\right)^2}\\
        &+& \theta\sum_{i=1}^n x_{ir}\Bigg[\frac{\left(\frac{t_i}{\exp(\mathbf{x}_i ^ \top \boldsymbol{\beta})}\right)^\alpha}{\left(1+\left(\frac{t_i}{\exp(\mathbf{x}_i ^ \top \boldsymbol{\beta})}\right)^\alpha\right)^2}+\theta\frac{
			\alpha\left(\frac{t_i}{\exp(\mathbf{x}_i ^ \top \boldsymbol{\beta})}\right)^\alpha
			\log\left(\frac{t_i}{\exp(\mathbf{x}_i ^ \top \boldsymbol{\beta})}\right)
			\left(1-\left(\frac{t_i}{\exp(\mathbf{x}_i ^ \top \boldsymbol{\beta})}\right)^\alpha\right)}{\left(1+\left(\frac{t_i}{\exp(\mathbf{x}_i ^ \top \boldsymbol{\beta})}\right)^\alpha\right)^3}
		\Bigg],
		\\
		\frac{\partial^2\ell\left(\bm{\psi}\right)}{\partial\alpha^2}&=&\sum_{i=1}^n
		\Bigg[-\frac{\delta_i}{\alpha^2}-\frac{2\delta_i
			\left(\frac{t_i}{\exp(\mathbf{x}_i ^ \top \boldsymbol{\beta})}\right)^\alpha
			\left(\log\left(\frac{t_i}{\exp(\mathbf{x}_i ^ \top \boldsymbol{\beta})}\right)\right)^2}
		{\left(1+\left(\frac{t_i}{\exp(\mathbf{x}_i ^ \top \boldsymbol{\beta})}\right)^\alpha\right)^2}\Bigg]
		\\
		&&-\theta\sum_{i=1}^n\frac{
			\left(\frac{t_i}{\exp(\mathbf{x}_i ^ \top \boldsymbol{\beta})}\right)^\alpha
			\left(\log\left(\frac{t_i}{\exp(\mathbf{x}_i ^ \top \boldsymbol{\beta})}\right)\right)^2
			\left(1-\left(\frac{t_i}{\exp(\mathbf{x}_i ^ \top \boldsymbol{\beta})}\right)^\alpha
			\right)}{\left(1+\left(\frac{t_i}{\exp(\mathbf{x}_i ^ \top \boldsymbol{\beta})}\right)^\alpha\right)^3}.
	\end{eqnarray*}

\noindent Similarly, the observed information matrix has the form, $\mathcal{H}\left(\bm{\psi}\right)=\left(-\frac{\partial^2 \ell\left(\bm{\psi}\right)}{\partial \bm{\psi}^\top \partial\bm{\psi}}\right)_{d\times d}$, with $u, b \in \{1, 2, \ldots, d\}$. The third-order cumulants, $\xi_{u,b,\nu}$, are derived using the same approach as outlined earlier.

\subsubsection{Third-order cumulants}

\begin{eqnarray*}
		\frac{\partial^3\ell\left(\bm{\psi}\right)}{\partial\theta^3}&=&
		2\sum_{i=1}^n\frac{\delta_i}{\theta^3}, \quad 
	\frac{\partial^3\ell\left(\bm{\psi}\right)}{\partial\theta^2\partial\beta_r}=0,\quad
		\frac{\partial^3\ell\left(\bm{\psi}\right)}{\partial\theta^2\partial\alpha}=0,\\
      \frac{\partial^3\ell\left(\bm{\psi}\right)}{\partial\theta\,\partial\beta_r\,\partial\beta_s}&=&
		-\sum_{i=1}^nx_{ir}x_{is}\,\frac{\alpha^2\left(\frac{t_i}{\exp(\mathbf{x}_i ^ \top \boldsymbol{\beta})}\right)^{\alpha}\left(1-\left(\frac{t_i}{\exp(\mathbf{x}_i ^ \top \boldsymbol{\beta})}\right)^{\alpha}\right)}{\left(1+\left(\frac{t_i}{\exp(\mathbf{x}_i ^ \top \boldsymbol{\beta})}\right)^{\alpha}\right)^{3}}
		\\
		\frac{\partial^3\ell\left(\bm{\psi}\right)}{\partial\theta\,\partial\beta_r\,\partial\alpha}
		&=&\sum_{i=1}^nx_{ir}\left[\frac{\left(\frac{t_i}
			{\exp(\mathbf{x}_i ^ \top \boldsymbol{\beta})}\right)^{\alpha}}{\left(1+\left(\frac{t_i}{\exp(\mathbf{x}_i ^ \top \boldsymbol{\beta})}\right)^{\alpha}\right)^{2}}+\frac{
			\alpha \left(\frac{t_i}{\exp(\mathbf{x}_i ^ \top \boldsymbol{\beta})}\right)^{\alpha}
			\left(1-\left(\frac{t_i}{\exp(\mathbf{x}_i ^ \top \boldsymbol{\beta})}\right)^{\alpha}
			\right) \log\left(\frac{t_i}{\exp(\mathbf{x}_i ^ \top \boldsymbol{\beta})}\right)}{\left(1+\left(\frac{t_i}{\exp(\mathbf{x}_i ^ \top \boldsymbol{\beta})}\right)^{\alpha}\right)^{3}}\right].
			\end{eqnarray*}

      \begin{eqnarray*}
		\frac{\partial^3\ell\left(\bm{\psi}\right)}{\partial\theta\,\partial\alpha^2}&=&
		-\sum_{i=1}^n\left(\frac{t_i}{\exp(\mathbf{x}_i ^ \top \boldsymbol{\beta})}\right)^{\alpha}
			\left(1-\left(\frac{t_i}{\exp(\mathbf{x}_i ^ \top \boldsymbol{\beta})}\right)^{\alpha}
			\right)\left(\log\left(\frac{t_i}{\exp(\mathbf{x}_i ^ \top \boldsymbol{\beta})}\right)\right)^2\left[1+\left(\frac{t_i}{\exp(\mathbf{x}_i ^ \top \boldsymbol{\beta})}\right)^{\alpha}\right]^{-3},
		\\
		\frac{\partial^3\ell\left(\bm{\psi}\right)}{\partial\beta_r\,\partial\beta_s\,\partial\beta_k}
		&=&
		2\alpha^3\sum_{i=1}^n
		x_{ir}x_{is}x_{ik}\delta_i
			\left(\frac{t_i}{\exp(\mathbf{x}_i ^ \top \boldsymbol{\beta})}\right)^{\alpha}
			\left(1-\left(\frac{t_i}{\exp(\mathbf{x}_i ^ \top \boldsymbol{\beta})}\right)^{\alpha}
			\right)\left[1+\left(\frac{t_i}{\exp(\mathbf{x}_i ^ \top \boldsymbol{\beta})}\right)^{\alpha}\right]^{-3}
		\\
		&+&
		\theta\alpha^3\sum_{i=1}^n
		x_{ir}x_{is}x_{ik}
		\Bigg[\frac{
			\left(\frac{t_i}{\exp(\mathbf{x}_i ^ \top \boldsymbol{\beta})}\right)^{\alpha}
			\left(1-4\left(\frac{t_i}{\exp(\mathbf{x}_i ^ \top \boldsymbol{\beta})}\right)^{\alpha}
			+\left(\left(\frac{t_i}{\exp(\mathbf{x}_i ^ \top \boldsymbol{\beta})}\right)^{\alpha}\right)^2
			\right)}{\left(1+\left(\frac{t_i}{\exp(\mathbf{x}_i ^ \top \boldsymbol{\beta})}\right)^{\alpha}\right)^{4}}
            \Bigg],
		\\
		\frac{\partial^3\ell\left(\bm{\psi}\right)}{\partial\alpha^3}&=&\sum_{i=1}^n
		\Bigg[\delta_i\left(\frac{2}{\alpha^3}-\frac{
			2\left(\log t_i-\mathbf{x}_i ^ \top \boldsymbol{\beta}\right)^3
			\left(\frac{t_i}{\exp(\mathbf{x}_i ^ \top \boldsymbol{\beta})}\right)^\alpha
			\left[1-\left(\frac{t_i}{\exp(\mathbf{x}_i ^ \top \boldsymbol{\beta})}\right)^\alpha
			\right]}{\left[1+\left(\frac{t_i}{\exp(\mathbf{x}_i ^ \top \boldsymbol{\beta})}\right)^\alpha
			\right]^3}\right)\Bigg]
		\\
		&&-\theta\,\sum_{i=1}^n
		\left[\frac{
			\left(\log t_i-\mathbf{x}_i ^ \top \boldsymbol{\beta}\right)^3
			\left(\frac{t_i}{\exp(\mathbf{x}_i^\top \boldsymbol{\beta})}\right)^\alpha
			\left[1-4\left(\frac{t_i}{\exp(\mathbf{x}_i ^ \top \boldsymbol{\beta})}\right)^\alpha
			+\left(\frac{t_i}{\exp(\mathbf{x}_i ^ \top \boldsymbol{\beta})}\right)^{2\alpha}
			\right]}{\left[1+\left(\frac{t_i}{\exp(\mathbf{x}_i ^ \top \boldsymbol{\beta})}\right)^\alpha
			\right]^4}\right].
	\end{eqnarray*}

\vspace{-0.3cm}
\begin{align*}
\begin{aligned}
\frac{\partial^3\ell\left(\bm{\psi}\right)}{\partial\beta_r\,\partial\beta_s\,\partial\alpha}
	=-\sum_{i=1}^nx_{ir}x_{is}
		\Bigg[
        &
		\frac{4\delta_i\alpha
			\left(\frac{t_i}{\exp(\mathbf{x}_i ^ \top \boldsymbol{\beta})}\right)^\alpha}
		{\left(1+\left(\frac{t_i}{\exp(\mathbf{x}_i^ \top \boldsymbol{\beta})}\right)^\alpha\right)^2}+
		\frac{2\delta_i\alpha^2
			\left(\frac{t_i}{\exp(\mathbf{x}_i ^ \top \boldsymbol{\beta})}\right)^\alpha
			\left(
			1-\left(\frac{t_i}{\exp(\mathbf{x}_i ^ \top \boldsymbol{\beta})}\right)^\alpha
			\right)
			\log\left(\frac{t_i}{\exp(\mathbf{x}_i ^ \top \boldsymbol{\beta})}\right)}{\left(1+\left(\frac{t_i}{\exp(\mathbf{x}_i ^ \top \boldsymbol{\beta})}\right)^\alpha\right)^3}\\
	       &+
           \theta
	\frac{2\alpha\left(\frac{t_i}{\exp(\mathbf{x}_i ^ \top \boldsymbol{\beta})}\right)^\alpha
		\left(1-\left(\frac{t_i}{\exp(\mathbf{x}_i^\top \boldsymbol{\beta})}\right)^\alpha\right)}{\left(1+\left(\frac{t_i}{\exp(\mathbf{x}_i^\top \boldsymbol{\beta})}\right)^\alpha\right)^3}\\
       & + \theta
\frac{ \alpha^2 \left(\frac{t_i}{\exp(\mathbf{x}_i ^ \top \boldsymbol{\beta})}\right)^\alpha
	\left(1-4\left(\frac{t_i}{\exp(\mathbf{x}_i ^ \top \boldsymbol{\beta})}\right)^\alpha
	+\left(\left(\frac{t_i}{\exp(\mathbf{x}_i ^ \top \boldsymbol{\beta})}\right)^\alpha\right)^2
	\right)}{\left(1+\left(\frac{t_i}{\exp(\mathbf{x}_i ^ \top \boldsymbol{\beta})}\right)^\alpha\right)^4}\times \frac{\log\left(\frac{t_i}{\exp(\mathbf{x}_i^\top \boldsymbol{\beta})}\right)}{\left(1+\left(\frac{t_i}{\exp(\mathbf{x}_i^\top \boldsymbol{\beta})}\right)^\alpha\right)^4}\Bigg]
\end{aligned}
\end{align*}

\vspace{-0.2cm}
\begin{align*}
	\begin{aligned}
		\frac{\partial^3\ell\left(\bm{\psi}\right)}{\partial\beta_r\,\partial\alpha^2}
		= \sum_{i=1}^n x_{ir}
		\Bigg[& \frac{4\delta_i
			\left(\frac{t_i}{\exp(\mathbf{x}_i ^ \top \boldsymbol{\beta})}\right)^\alpha
			\log\left(\frac{t_i}{\exp(\mathbf{x}_i ^ \top \boldsymbol{\beta})}\right)}{\left(1+\left(\frac{t_i}{\exp(\mathbf{x}_i ^ \top \boldsymbol{\beta})}\right)^\alpha\right)^2}
		\\
		&+\frac{ 2\delta_i\alpha
			\left(\frac{t_i}{\exp(\mathbf{x}_i ^ \top \boldsymbol{\beta})}\right)^\alpha
			\left(1-\left(\frac{t_i}{\exp(\mathbf{x}_i ^ \top \boldsymbol{\beta})}\right)^\alpha
			\right)\left(\log\left(\frac{t_i}{\exp(\mathbf{x}_i ^ \top \boldsymbol{\beta})}\right)\right)^2
		}{\left(1+\left(\frac{t_i}{\exp(\mathbf{x}_i ^ \top \boldsymbol{\beta})}\right)^\alpha\right)^3
	}\\
	&+\frac{2\theta\left(\frac{t_i}{\exp(\mathbf{x}_i ^ \top \boldsymbol{\beta})}\right)^\alpha
		\left(1-\left(\frac{t_i}{\exp(\mathbf{x}_i ^ \top \boldsymbol{\beta})}\right)^\alpha
		\right)\log\left(\frac{t_i}{\exp(\mathbf{x}_i ^ \top \boldsymbol{\beta})}\right)}{\left(1+\left(\frac{t_i}{\exp(\mathbf{x}_i ^ \top \boldsymbol{\beta})}\right)^\alpha\right)^3}\\
	&+
	\frac{\theta\alpha\left(\frac{t_i}{\exp(\mathbf{x}_i ^ \top \boldsymbol{\beta})}\right)^\alpha
		\left(1-4\left(\frac{t_i}{\exp(\mathbf{x}_i ^ \top \boldsymbol{\beta})}\right)^\alpha
		+\left(\frac{t_i}{\exp(\mathbf{x}_i ^ \top \boldsymbol{\beta})}\right)^{2\alpha}
		\right)\left(\log\left(\frac{t_i}{\exp(\mathbf{x}_i ^ \top \boldsymbol{\beta})}\right)\right)^2
	}{\left(1+\left(\frac{t_i}{\exp(\mathbf{x}_i ^ \top \boldsymbol{\beta})}\right)^\alpha\right)^4}
	\Bigg]. 
\end{aligned}
\end{align*}

\newpage
\section{Additional Monte Carlo simulation results}
\label{apendiceA2}
This second section of the appendix presents additional results that complement and reinforce the findings from the analysis of the synthetic data in Section~\ref{sec3}.

\begin{figure}[H]
$$
\begin{array}{cc}
 \includegraphics[scale=0.2]{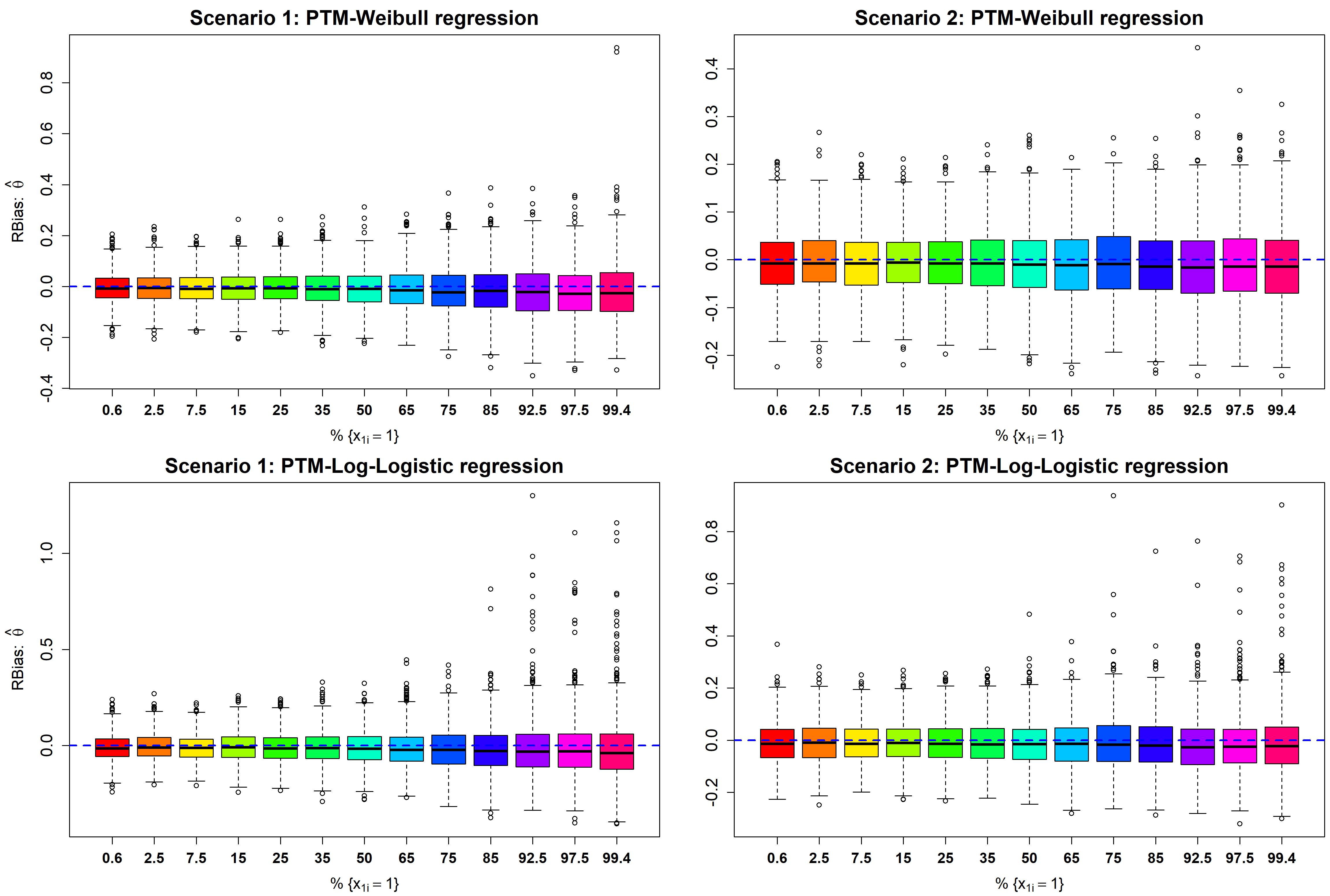}& \includegraphics[scale=0.2]{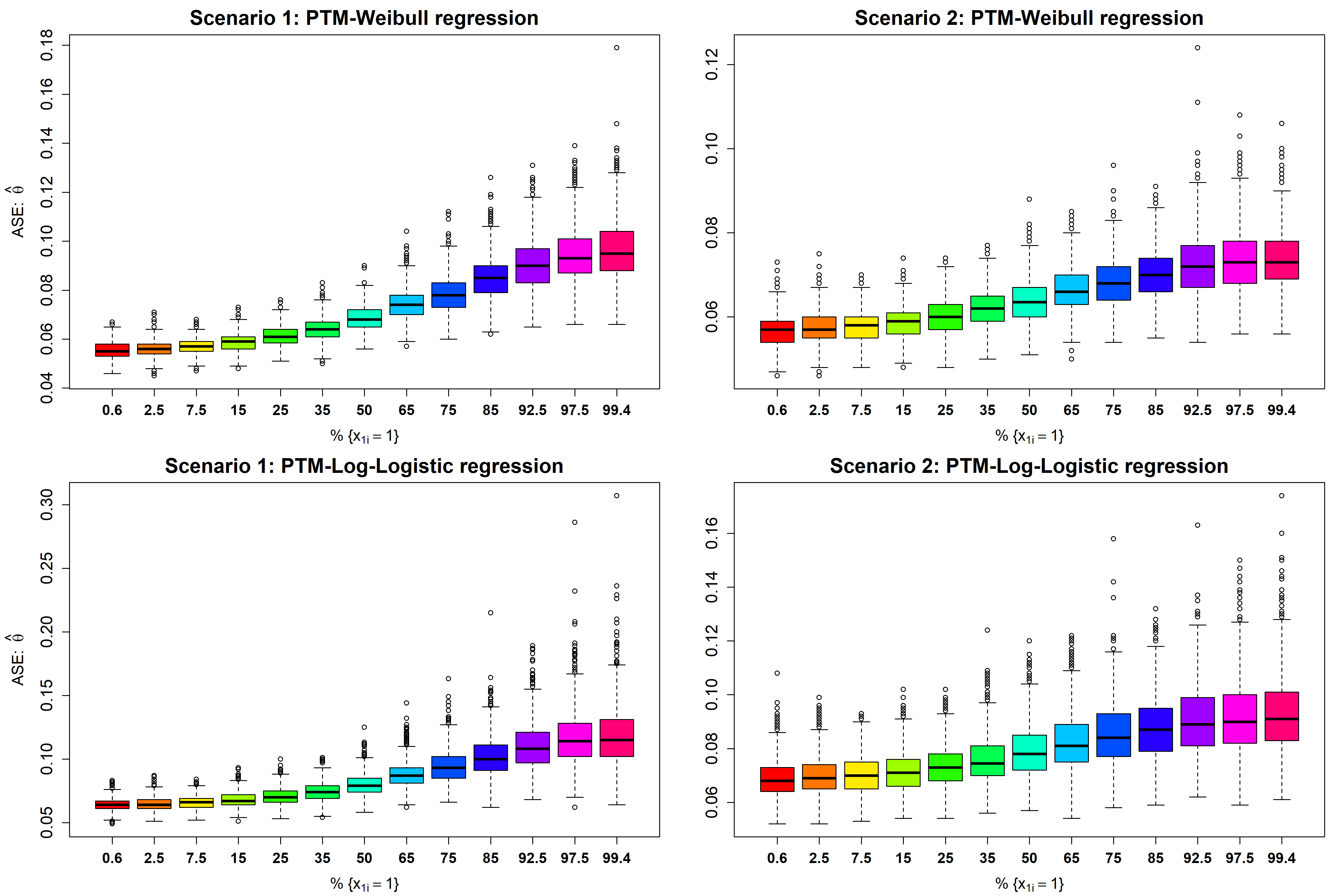}\\
 \includegraphics[scale=0.2]{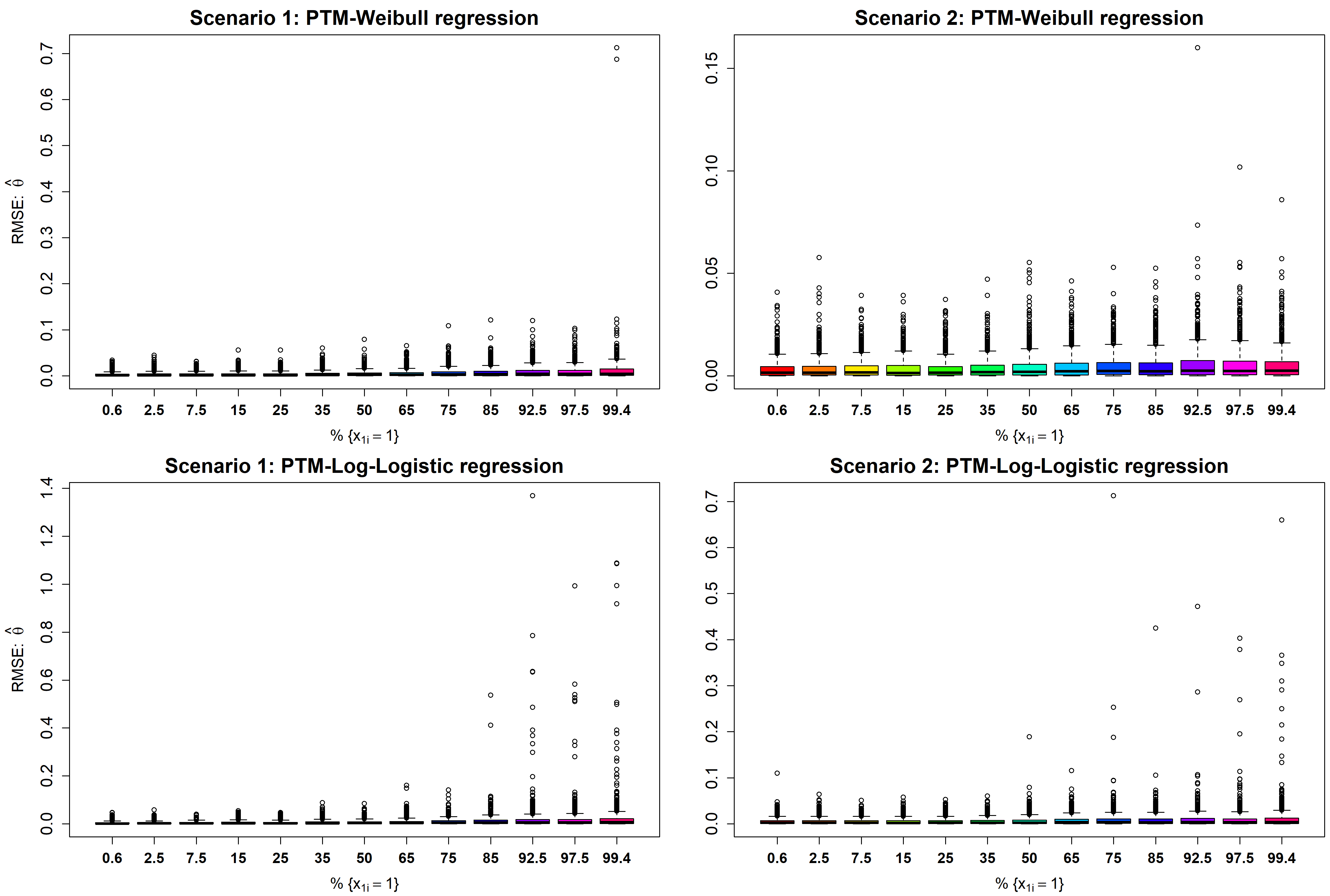}& \includegraphics[scale=0.2]{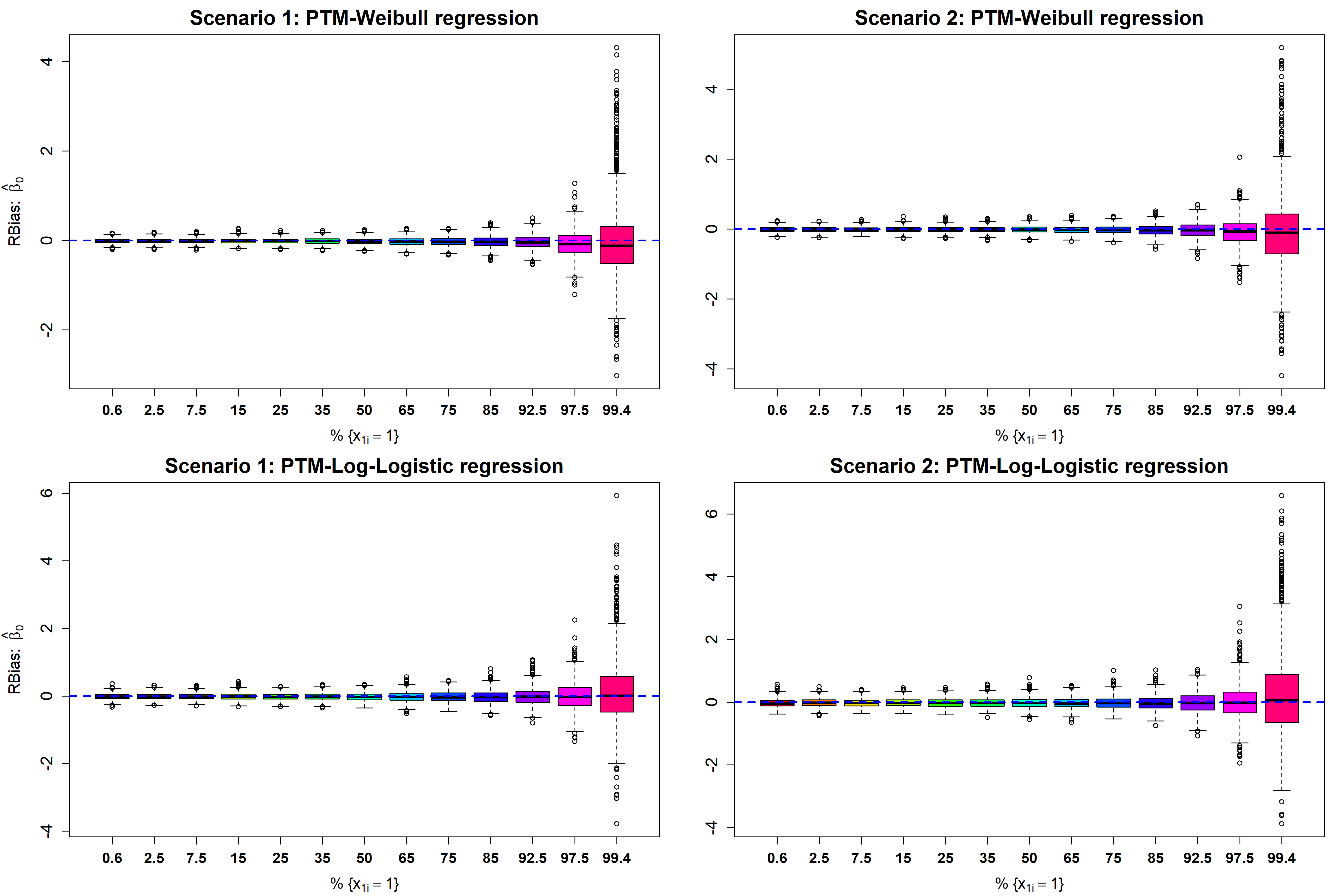}\\
 \includegraphics[scale=0.2]{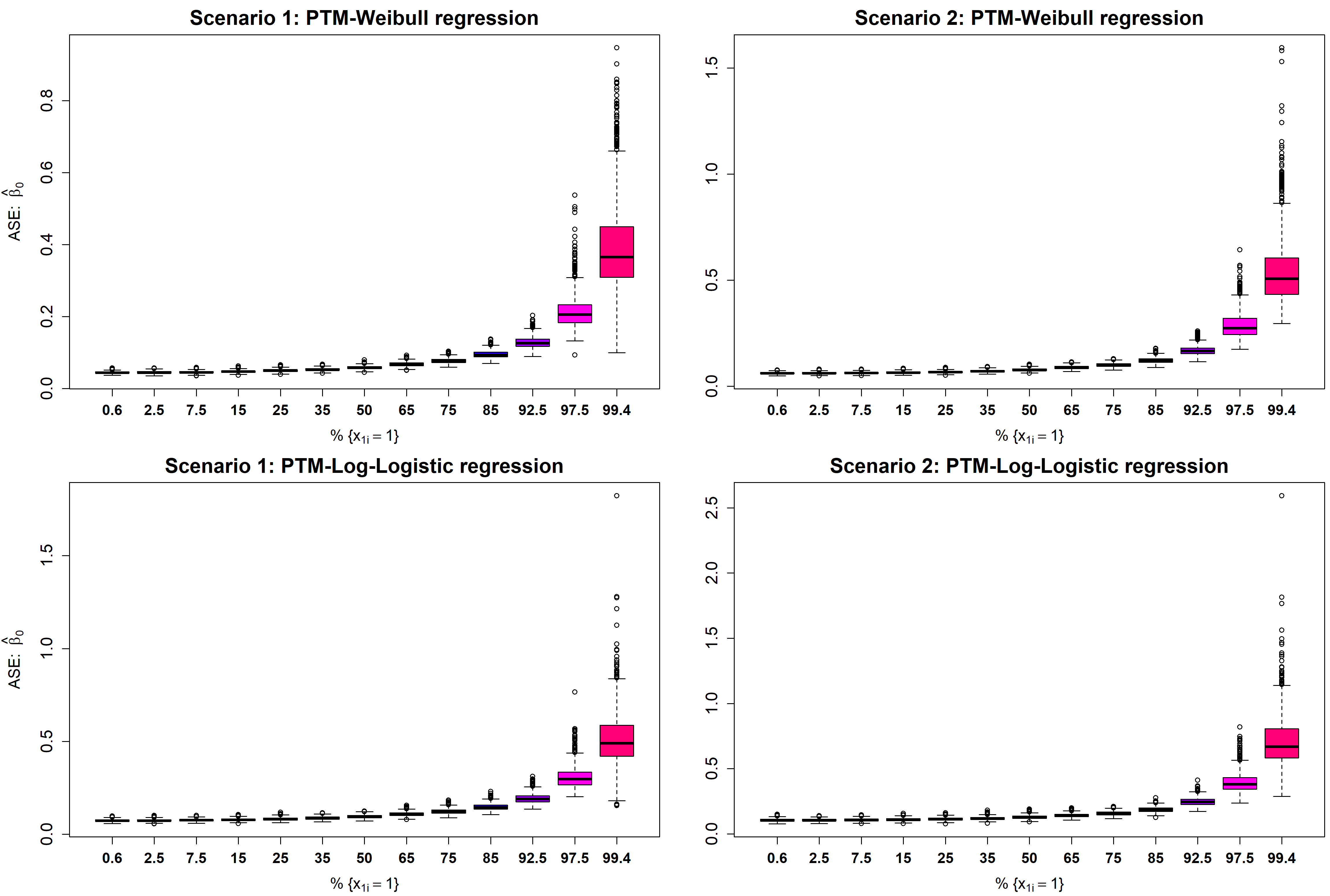} & \includegraphics[scale=0.2]{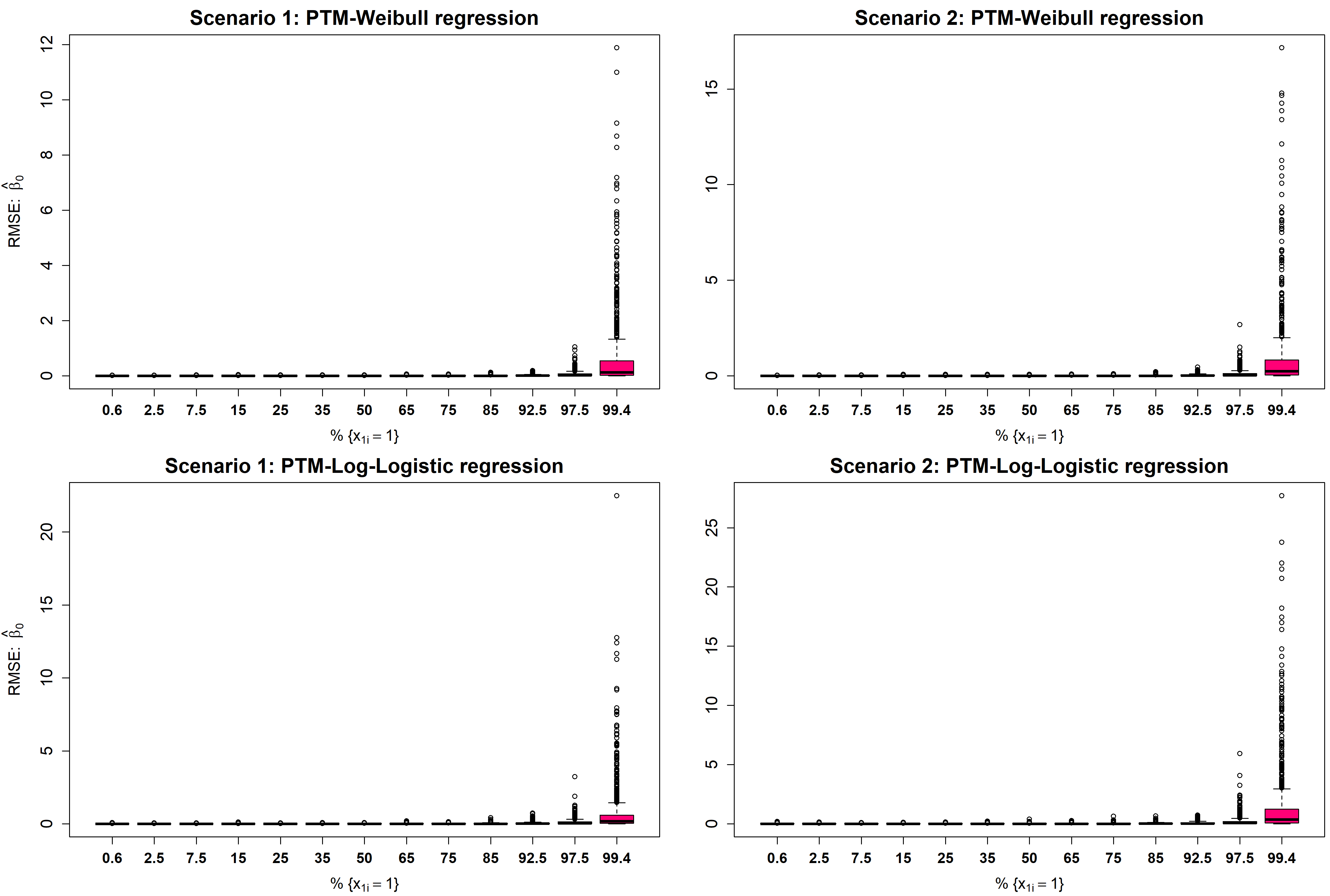}\\
\end{array}
$$
\vspace{-0.8cm}
\caption{Main performance measures for the estimates $\theta$ and $\beta_0$ under different proportions of successes in the binary covariate, $\%\{x_{1i} = 1\}$, with $n = 1000$ fixed.}
\label{Perform4}
\end{figure}

\newpage
\begin{figure}[H]
$$
\begin{array}{cc}
 \includegraphics[scale=0.2]{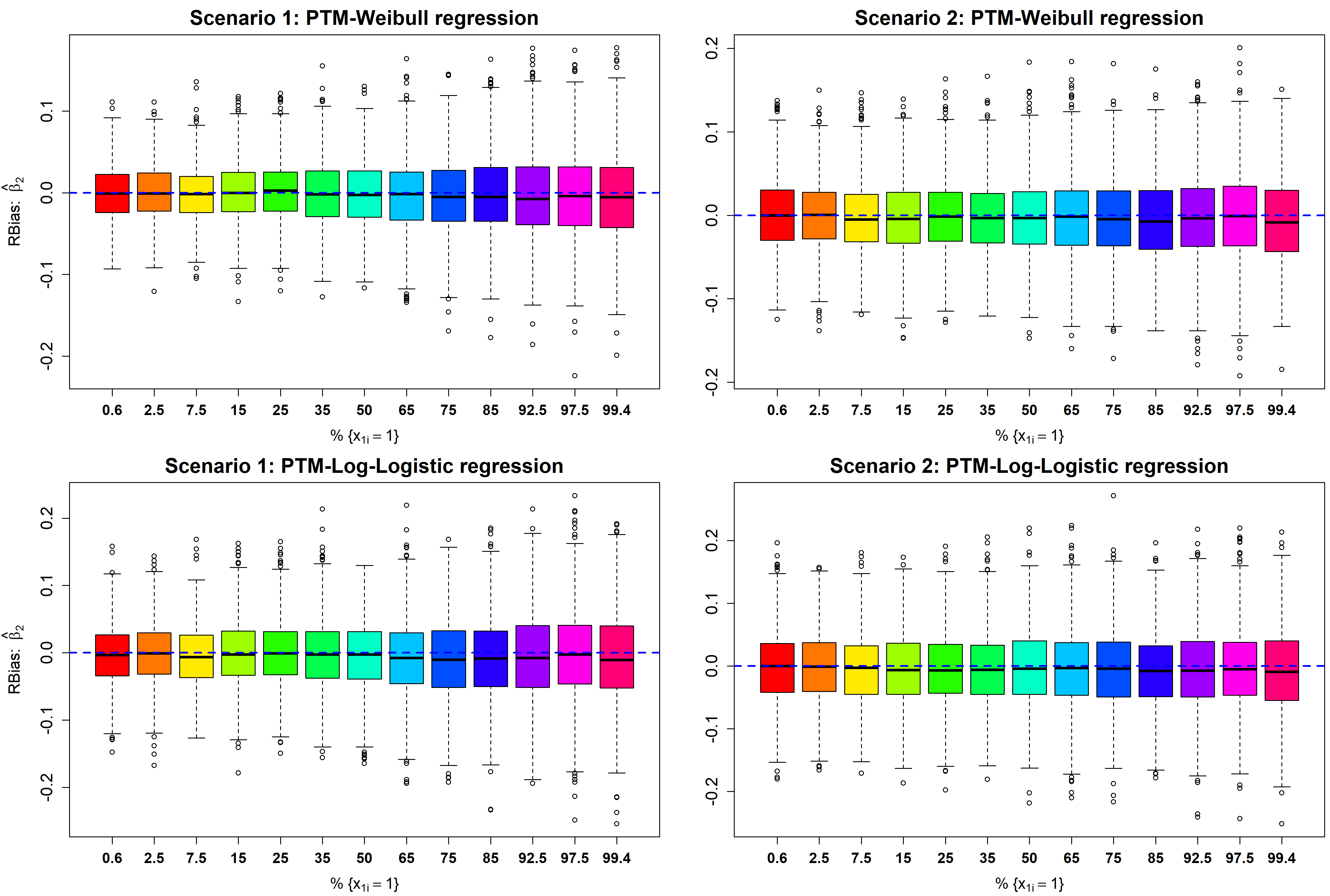}& \includegraphics[scale=0.2]{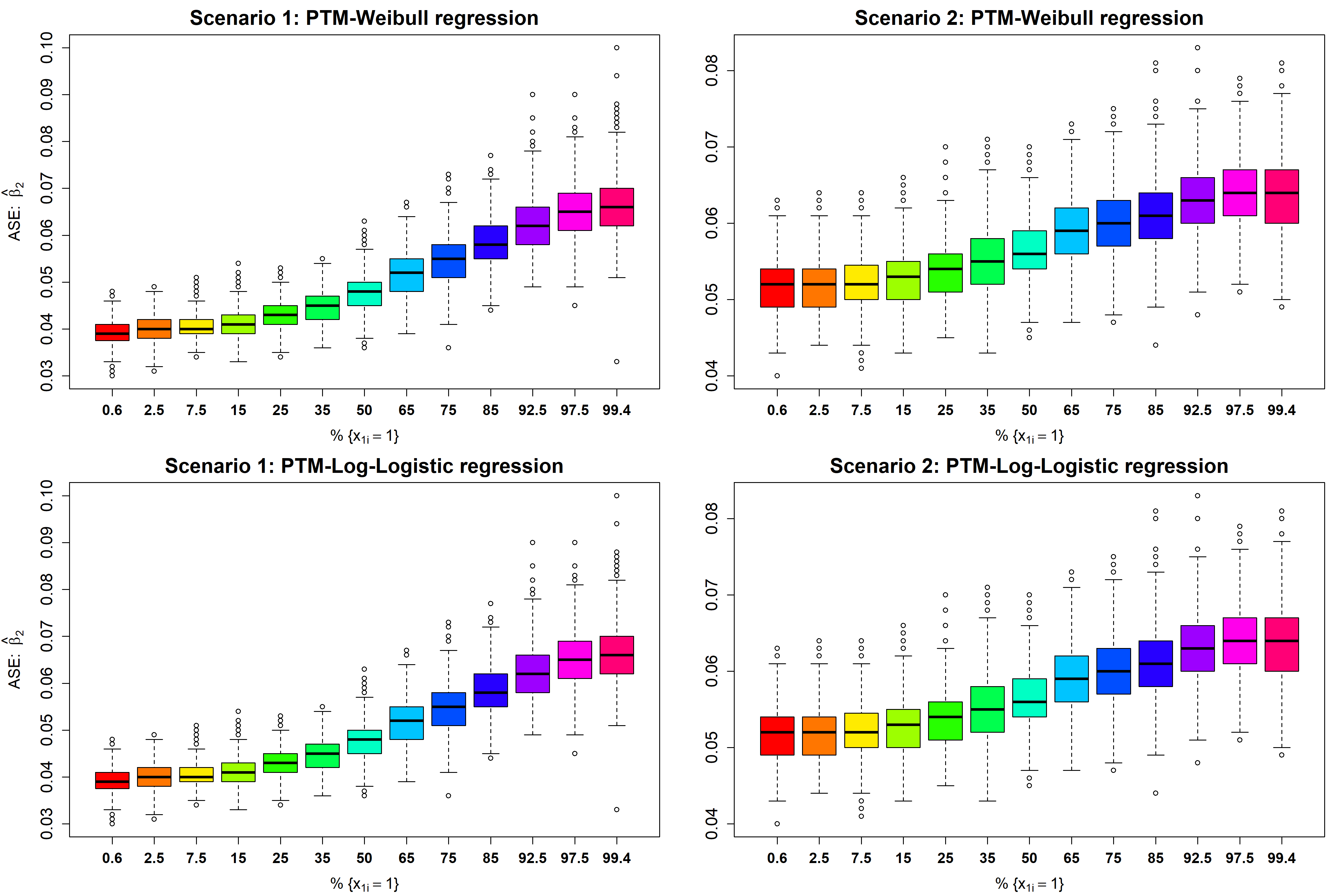}\\
 \includegraphics[scale=0.2]{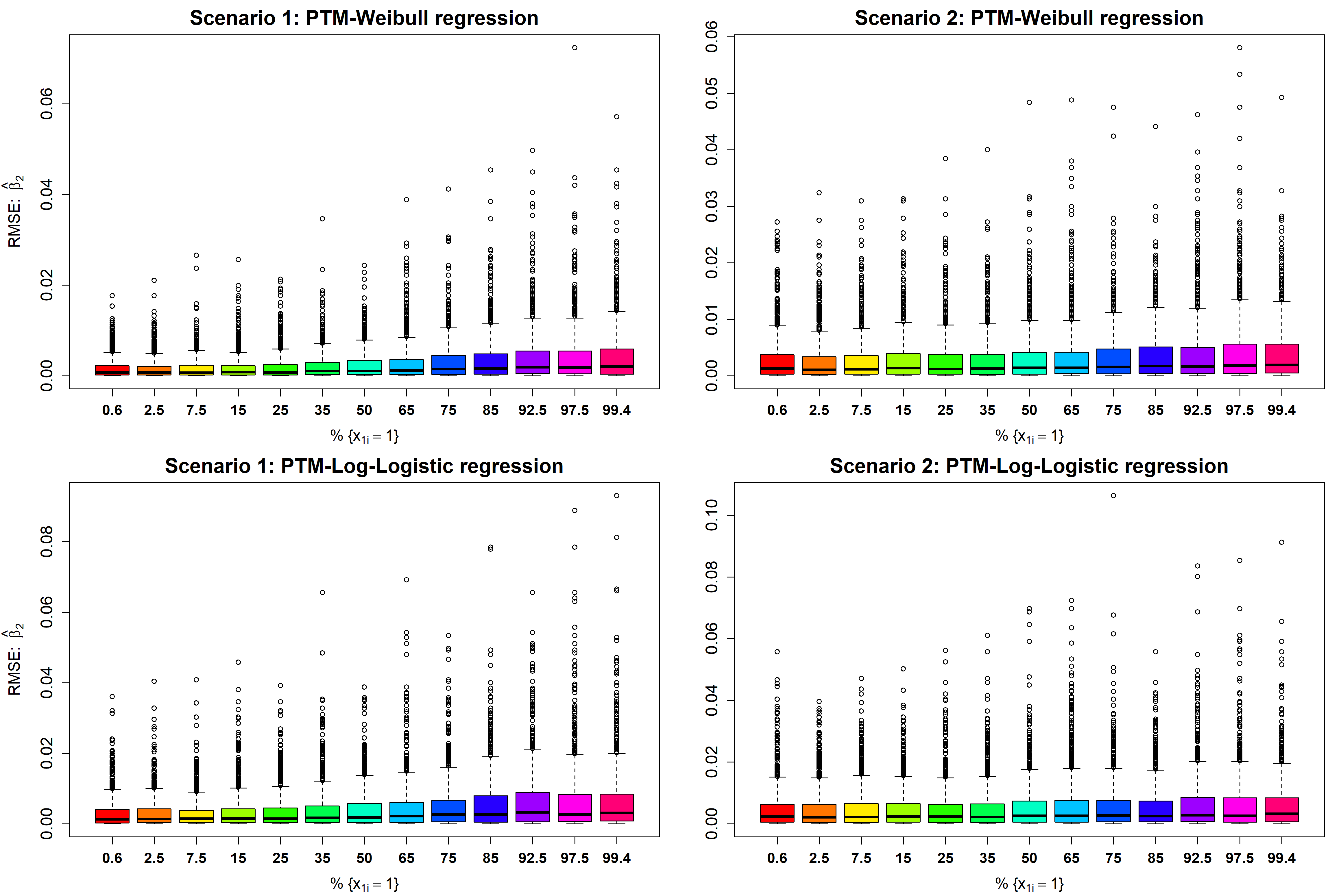}& \includegraphics[scale=0.2]{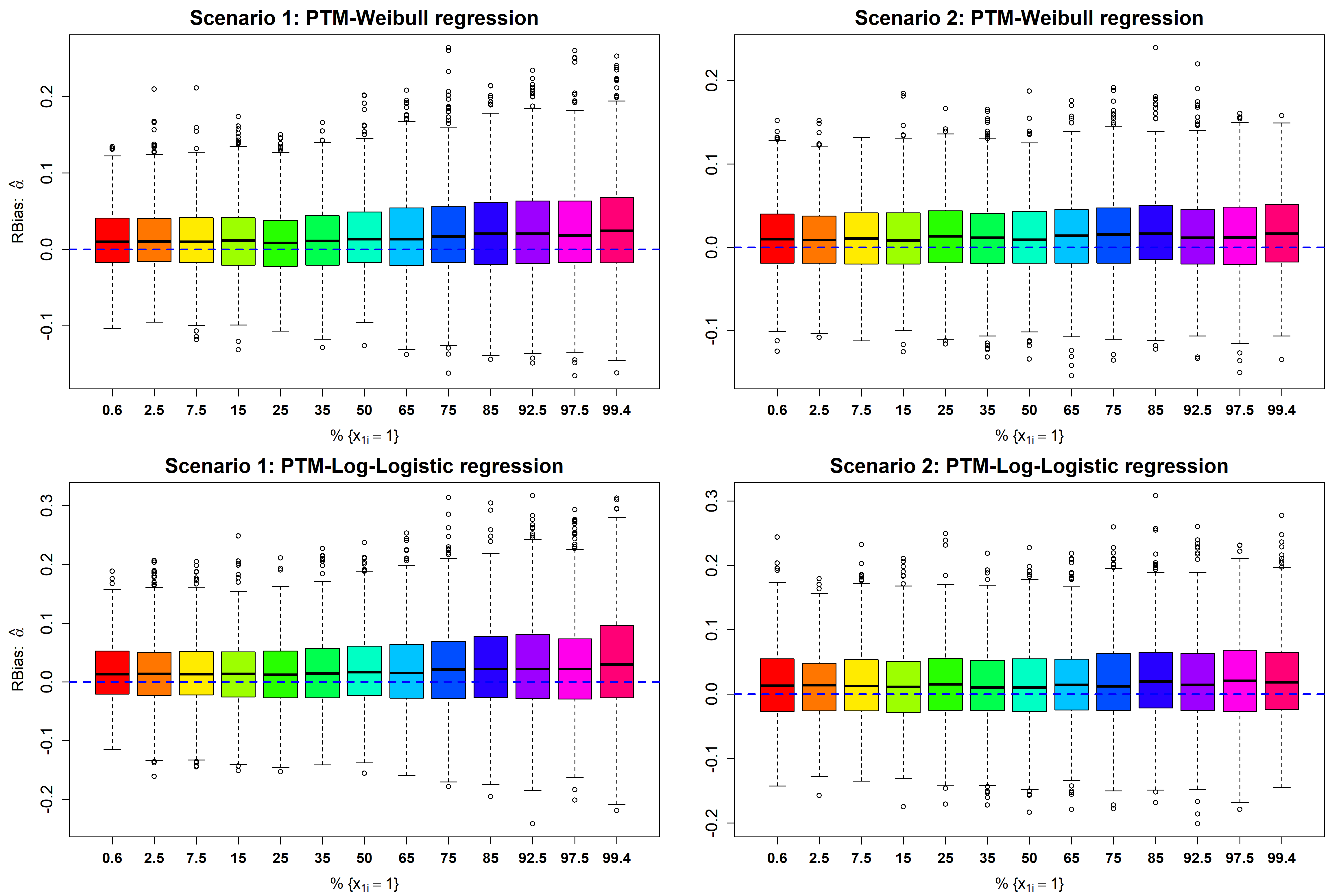}\\
 \includegraphics[scale=0.2]{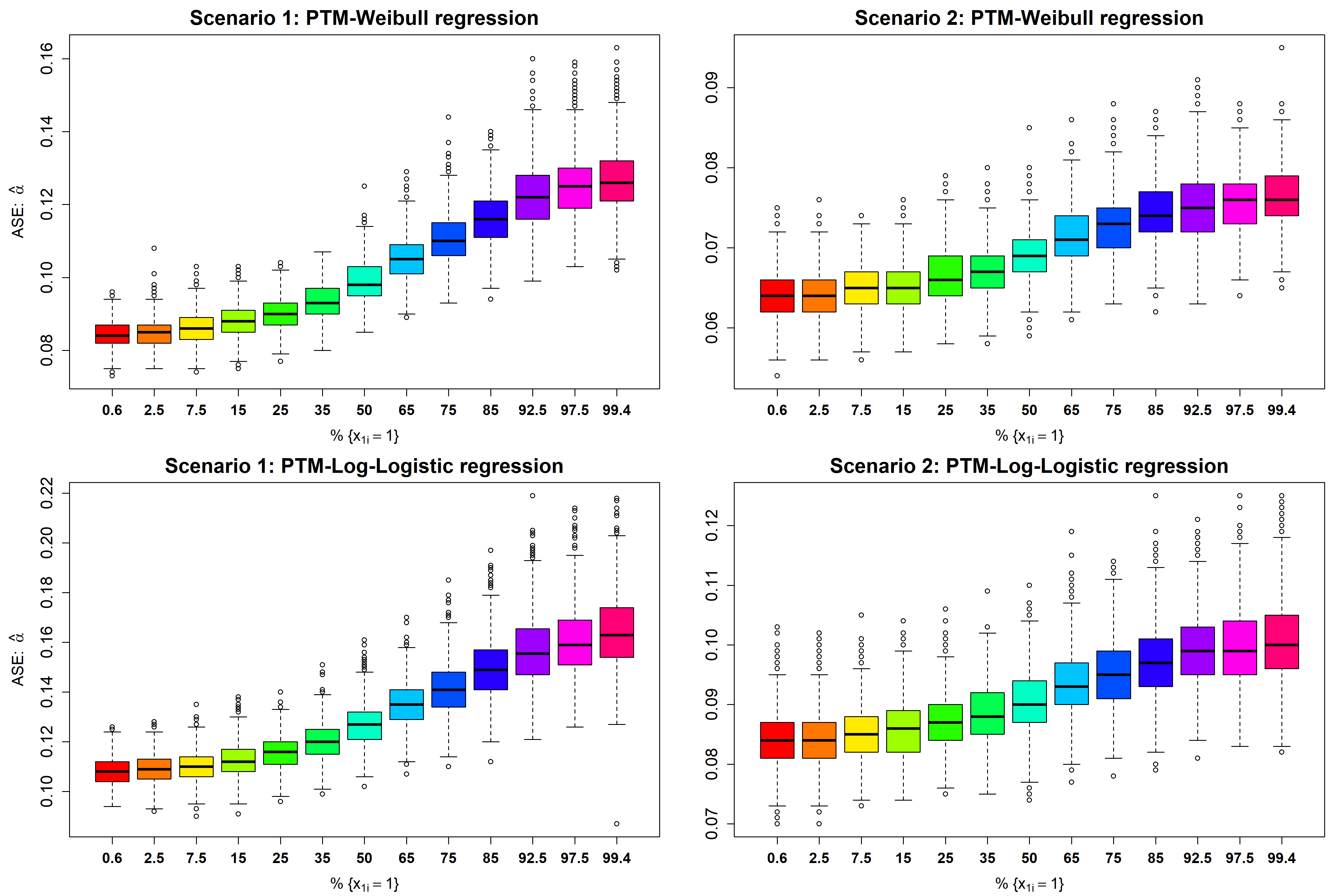} & \includegraphics[scale=0.2]{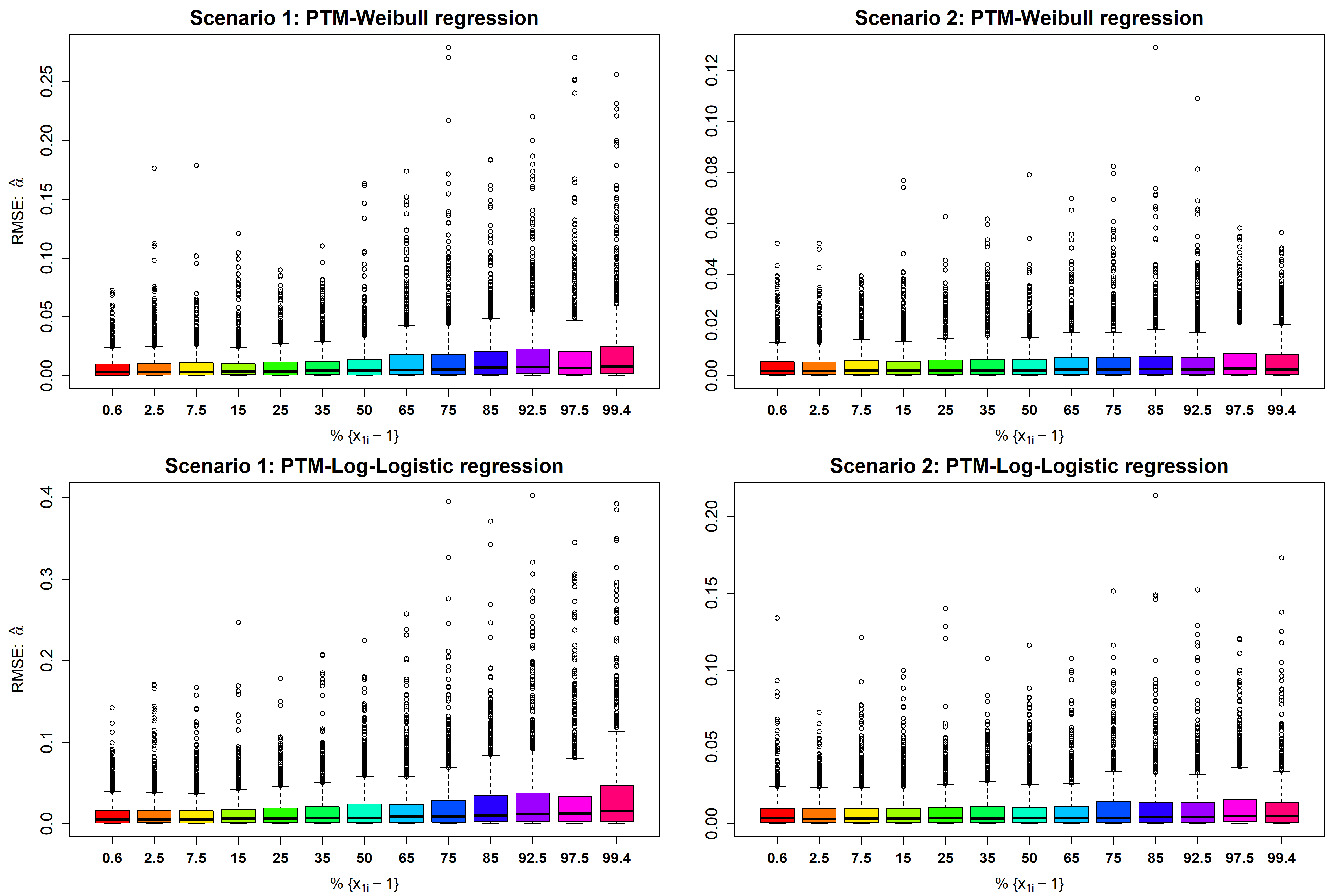}\\
\end{array}
$$
\vspace{-0.8cm}
\caption{
Main performance measures for the estimates $\hat{\beta_2}$ and $\hat{\alpha}$ under different proportions of successes in the binary covariate, $\%\{x_{1i}=1\}$, with $n = 1000$ fixed.}
\label{Perform3}
\end{figure}

\begin{landscape}
\begin{figure}[H]
	$$
	\begin{array}{cc}
	\mbox{Panel (a)}& \mbox{Panel (b)}\\
    \includegraphics[scale=0.32]{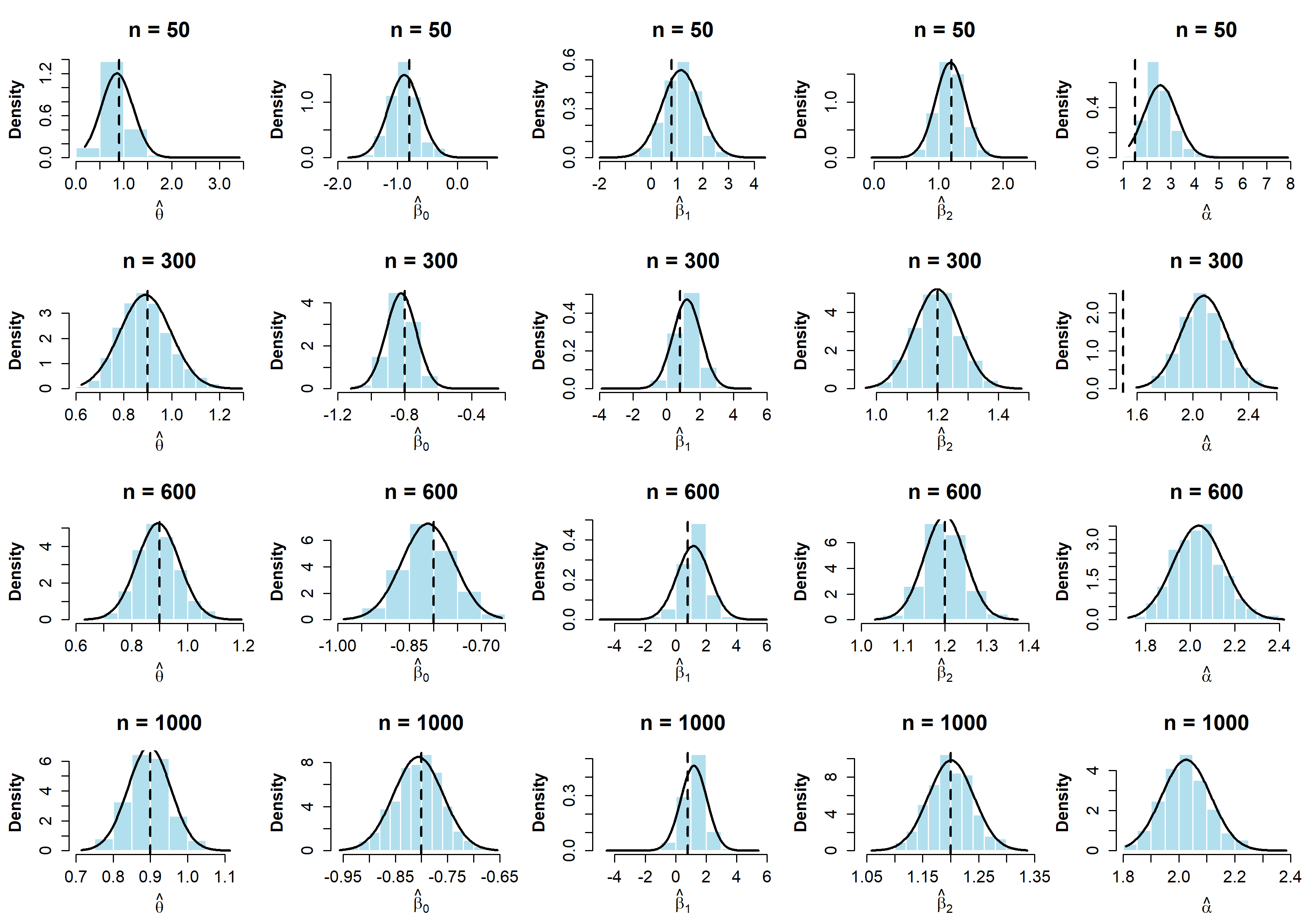}& \includegraphics[scale=0.32]{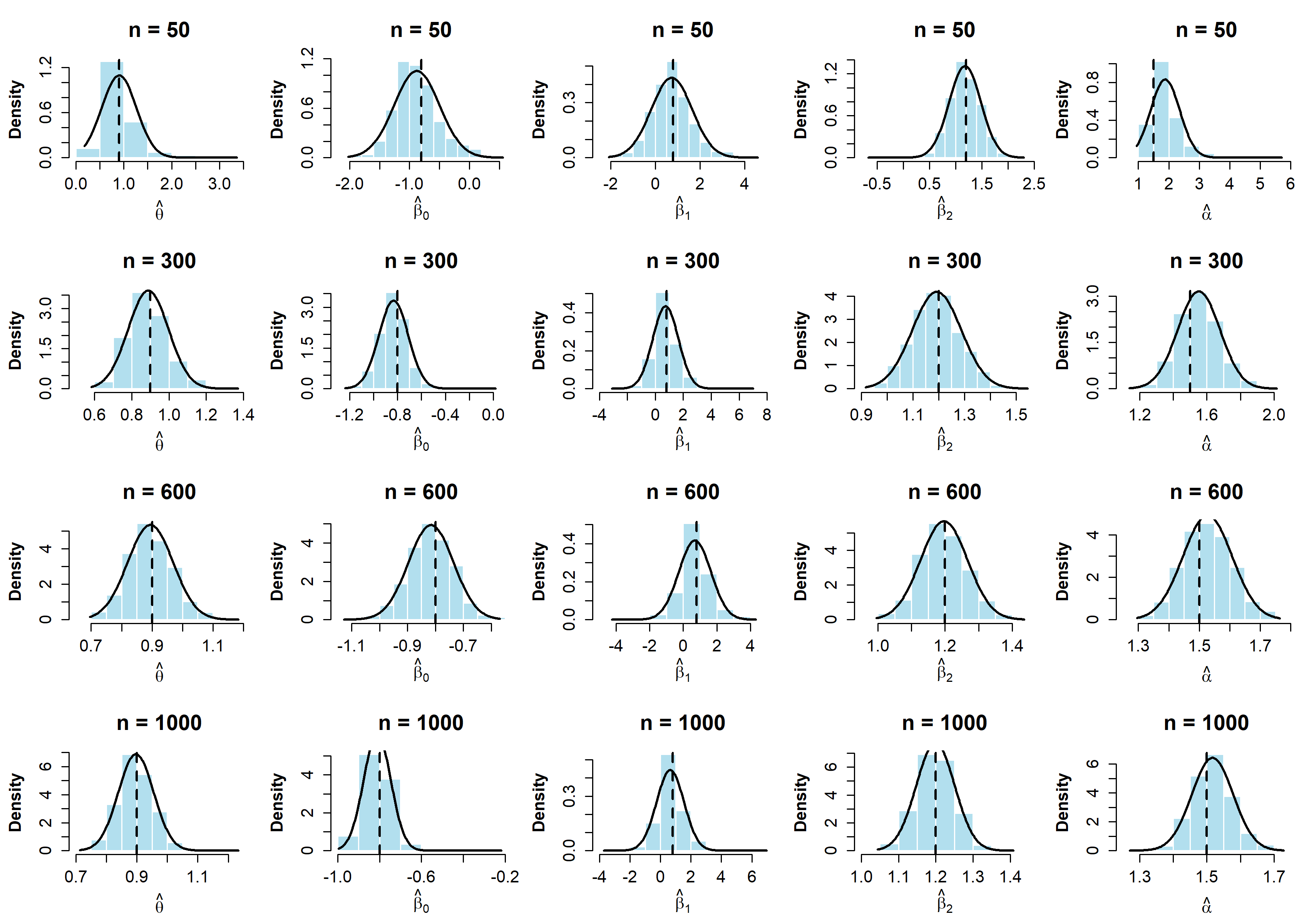}\\
    \mbox{Panel (c)}& \mbox{Panel (d)}\\
	\includegraphics[scale=0.32]{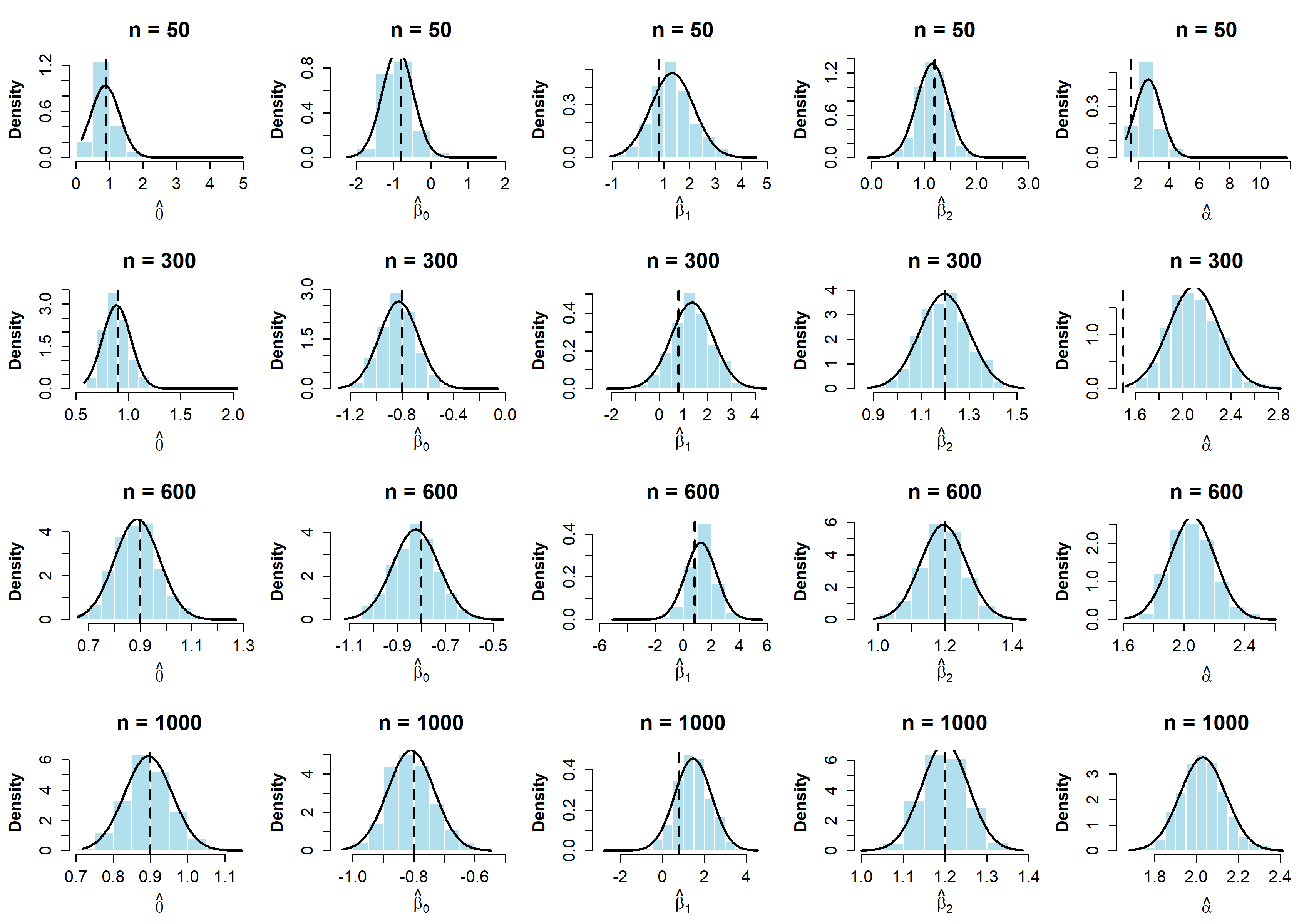}& \includegraphics[scale=0.32]{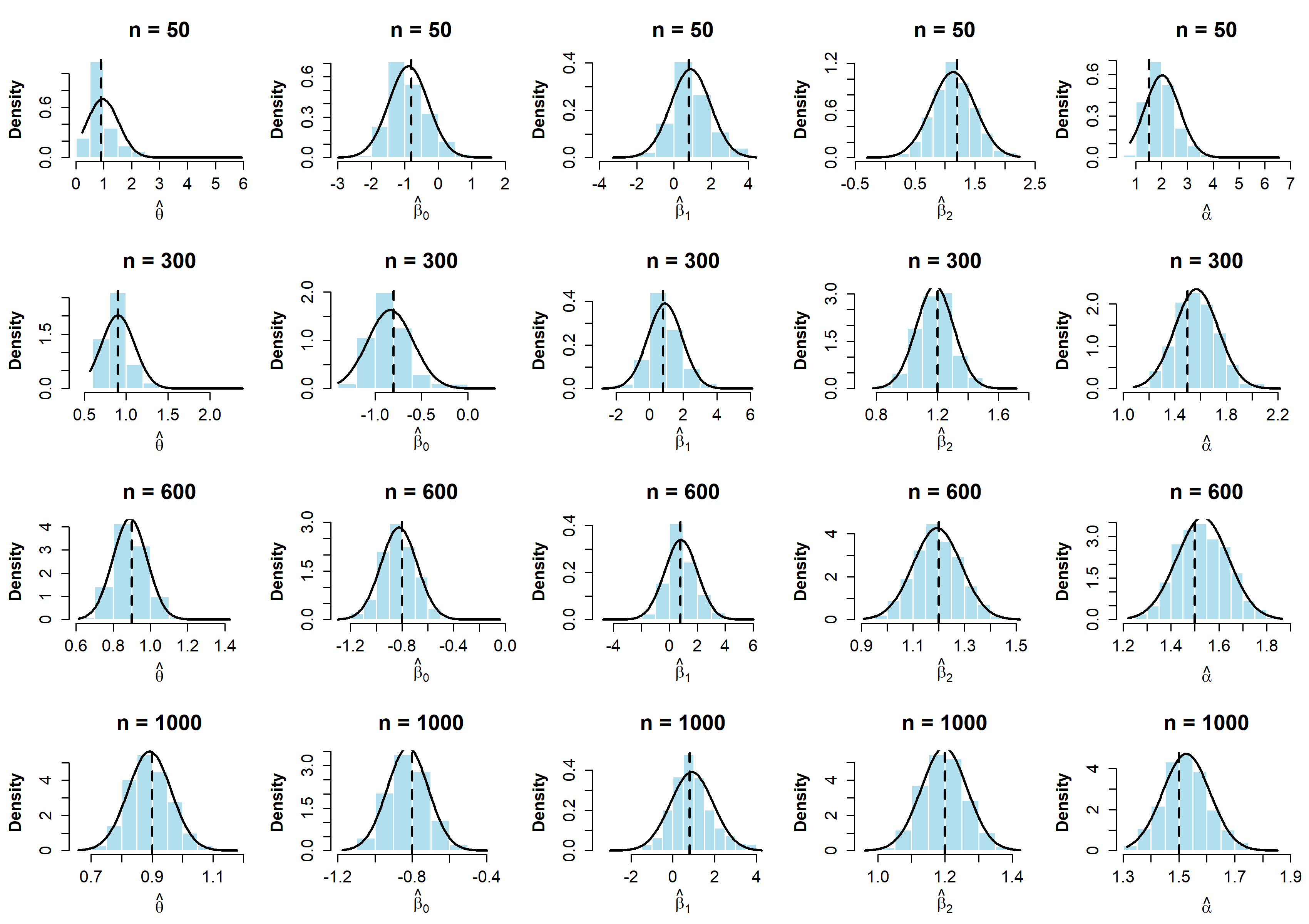}\\
	\end{array}
	$$
	\vspace{-0.8cm}
	\caption{Histograms illustrating the distributions of the Monte Carlo estimates for the quantities of interest. Panels $(a)$ and $(b)$ present the results for the PTM-Weibull regression model under Scenarios 1 and 2, respectively, while panels $(c)$ and $(d)$ display the results for the PTM-Log-Logistic model under Scenarios 1 and 2, respectively.}
\label{Perform3}
\end{figure}
\end{landscape}

\end{document}